%% file: main.tex
\documentclass[sigplan,screen]{acmart}
\AtBeginDocument{%
  }

\copyrightyear{2027}
\acmYear{2027}
\setcopyright{cc}
\setcctype{by}
\acmConference[ASPLOS '27]{Proceedings of the 32nd ACM International Conference on Architectural Support for Programming Languages and Operating Systems, Volume 1}{April 11--15, 2027}{Heraklion, Greece}
\acmBooktitle{Proceedings of the 32nd ACM International Conference on Architectural Support for Programming Languages and Operating Systems, Volume 1 (ASPLOS '27), April 11--15, 2027, Heraklion, Greece}
\acmDOI{10.1145/3845814.3850085}
\acmISBN{979-8-4007-3012-2/2027/04}

\usepackage{pifont}
\usepackage{comment}
\usepackage{multirow}
\usepackage{threeparttable}
\usepackage{enumitem}
\usepackage{xurl}

\usepackage{array}
\usepackage{booktabs}
\usepackage{tabularx}

\newcommand{\cmttfont}[1]{{\fontfamily{cmtt}\selectfont #1}}

\usepackage{listings,listings-rust}
\usepackage{siunitx}
\usepackage{balance}
\begin{document}

\settopmatter{printfolios=false,printacmref=true}
\title[safeL4]{safeL4: Rethinking Safe-Language-Based Single Address Space Microkernels}

%
\author{Zheyuan Xu}
\affiliation{%
  \department{Key Laboratory of Cyberspace Security Defense}
  \institution{Institute of Information Engineering, Chinese Academy of Sciences}
    \city{Beijing}
  \country{China}
}

\affiliation{%
    \department{School of Cyberspace Security}
  \institution{University of Chinese Academy of Sciences}
  \city{Beijing}
  \country{China}
}

\email{xuzheyuan@iie.ac.cn}

\author{Xiaobei Zhang}
\affiliation{%
  \department{Key Laboratory of Cyberspace Security Defense}
  \institution{Institute of Information Engineering, Chinese Academy of Sciences}
    \city{Beijing}
  \country{China}}
\email{zhangxiaobei@iie.ac.cn}

\author{Zihan Li}
\affiliation{%
  \department{Key Laboratory of Cyberspace Security Defense}
  \institution{Institute of Information Engineering, Chinese Academy of Sciences}
    \city{Beijing}
  \country{China}}
\email{zihanli0822@gmail.com}

\author{Zekai Nie}
\affiliation{%
  \department{Key Laboratory of Cyberspace Security Defense}
  \institution{Institute of Information Engineering, Chinese Academy of Sciences}
    \city{Beijing}
  \country{China}}
\email{niezekai@iie.ac.cn}

\author{Guangtao Zhu}
\affiliation{%
  \institution{UNSW Sydney}
  \city{Sydney}
  \state{NSW}
  \country{Australia}}
\email{guangtao.zhu@student.unsw.edu.au}

\author{Zhuang Lu}
\affiliation{%
  \department{Key Laboratory of Cyberspace Security Defense}
  \institution{Institute of Information Engineering, Chinese Academy of Sciences}
  \city{Beijing}
  \country{China}}
\email{luzhuang@iie.ac.cn}

\author{Qianying Zhang}
\affiliation{%
  \institution{Capital Normal University}
    \city{Beijing}
  \country{China}}
\email{qyzhang@cnu.edu.cn}

\author{Shijun Zhao}
\authornote{Corresponding author}
\affiliation{%
  \department{Key Laboratory of Cyberspace Security Defense}
  \institution{Institute of Information Engineering, Chinese Academy of Sciences}
    \city{Beijing}
  \country{China}}
\email{zhaoshijun@iie.ac.cn}

\author{Haibo Chen}
\affiliation{%
  \institution{Shanghai Jiao Tong University}
    \city{Shanghai}
  \country{China}}
\email{haibochen@sjtu.edu.cn}

\author{Rui Hou}
\affiliation{%
  \department{Key Laboratory of Cyberspace Security Defense}
  \institution{Institute of Information Engineering, Chinese Academy of Sciences}
    \city{Beijing}
  \country{China}}
\email{hourui@iie.ac.cn}

\renewcommand{\shortauthors}{Zheyuan Xu et al.}


\begin{abstract}
Microkernels provide good features such as high security, safety, and flexibility, but their reliance on hardware-based isolation leads to heavy performance overhead, limiting their widespread adoption. Although constructing single-address-space microkernels based on safe language-based isolation is a promising approach to addressing the performance issue, this approach still faces challenges in maintaining the principle of minimality and enforcing capability-based, fine-grained access control over system resources.
We propose safeL4, an approach to constructing microkernels based on safe languages. It guarantees minimality through low-level abstractions for unsafe operations, and implements a capability system based on the safe language's object capabilities to achieve fine-grained access control over all operations on system resources. We implement a prototype system in Rust, and our evaluation results show that the primitives provided by safeL4 are, on average, 1.9$\times$ faster than those of seL4.

  
\end{abstract}



\maketitle

\input{sections/1.introduction}
\input{sections/2.background}
\input{sections/3.motivation}

\input{sections/4.design}
\input{sections/5.implementation}
\input{sections/6.evaluation}
\input{sections/8.conclusion}
\begin{acks}
We would like to thank the reviewers for their comments, and our anonymous shepherd for their insightful comments and suggestions that greatly inspired us.
We also thank Gernot Heiser for his insightful comments on safeL4.
This work is supported by the National Natural Science Foundation of China (No.62272458 and No.62372311), and the National Natural Science Foundation of China for Distinguished Young Scholars (No.62125208).
\end{acks}

\appendix
\input{sections/safeL4_ae_appendix}

\bibliographystyle{ACM-Reference-Format}
\balance
\bibliography{sample}

\end{document}

%% file: sections/1.introduction.tex
\section{Introduction}
Despite performance concerns, microkernels have long been regarded as offering significant advantages over monolithic kernels~\cite{liedtke1996toward,herder2005towards}, primarily in security, safety, and flexibility. First, microkernels adhere to the principle of minimality~\cite{lied95}, ensuring that the \emph{trusted computing base} (TCB) is as small as possible. Second, microkernels utilize capability systems~\cite{bomberger1992keykos,lackorzynski2009taming,shapiro1999eros} to implement fine-grained access control over system resources, thereby enforcing the \emph{principle of least privilege} (PoLP)~\cite{polp,jero2021practical}. The minimality principle and capability-based access control enable formal verification of the kernel~\cite{sel42013}, leading to the world’s first formally verified \emph{operating system} (OS) kernel, seL4~\cite{murray2013sel4,sewell2011sel4,klein2014comprehensive}. Finally, by partitioning OS services into domains that can be updated or replaced independently, microkernels achieve high flexibility. In the remainder of this paper, we refer to the domains protected by the microkernel as \emph{protection domains}~\cite{lampson1974protection}.

Microkernels, due to their advantages, were once envisioned as the basis for all types of operating systems~\cite{liedtke1996toward}. However, since the first generation of microkernels~\cite{accetta1986mach}, they have been used mainly in embedded and safety-critical environments (i.e., essentially static systems). In contrast, monolithic kernels remain the most prevalent and widely adopted kernel architecture. Emerging scenarios, such as smart vehicles, smartphones, trusted execution environments, containers, and confidential virtual machines, have strict security and safety requirements and therefore favor microkernels~\cite{hm2024,weinhold2023towards,miemietz2024perfect,miemietz2025metteagle,shi2025serverless}. These scenarios are no longer limited to embedded systems and involve running general-purpose applications that demand a broad range of OS functionality.

Nevertheless, the hardware-based isolation used by microkernels imposes significant performance overhead in these scenarios, including context switches caused by \emph{inter-process communication} (IPC), cross-address-space data transfers~\cite{du2019xpc}, and increased IPC frequency~\cite{hm2024}. Although various measures have been proposed to mitigate the performance issues of microkernels, none has fundamentally solved the problem. For example, some OS designs~\cite{bricker1991new,hm2024} place the most critical servers and drivers in the kernel, which conflicts with the principle of minimality; moreover, some measures duplicate kernel state across multiple servers~\cite{hm2024}, introducing extra overhead for state synchronization.


In fact, there is another technical approach that aims to design microkernels that do not rely on hardware-based isolation to eliminate the performance overhead, namely, the use of type-safe languages. 
These languages use type systems to define clear contracts for components (such as modules, structures, and functions), ensuring that each component interacts with others only through well-defined interfaces and preventing unauthorized access~\cite{li2021understanding,burtsev2021isolation}.
Safe language-based microkernels and applications run in the same address space and the same CPU mode, i.e., a \emph{single-address-space OS} (SASOS), but they still follow the common concept of a kernel, namely, ``the part of the operating system that is mandatory and common to all other software''~\cite{lied95}. 
Such a language-based approach significantly reduces the performance overhead introduced by hardware isolation: cross-domain communication is simplified to function calls, and cross-domain data transfer is achieved via zero-copy object ownership transfer. 
Typical representatives of safe language-based microkernels include SPIN~\cite{bershad1994spin,spin1995}, Singularity~\cite{hunt2005overview,fahndrich2006language,singularity07}, and RedLeaf~\cite{redleaf20}. 
There are also some kernels that, although not explicitly claiming to be microkernels, have adopted certain microkernel features in their design, such as minimizing the TCB (Asterinas~\cite{peng2025asterinas}) and fault isolation (Tock~\cite{tock17,tock25}, Theseus~\cite{theseus20}).

However, current safe language-based kernels 
face two main issues in meeting microkernel requirements. 
First, these kernels lack a capability system for fine-grained access control over all system resources. 
A capability system is the principal mechanism microkernels use to restrict the system resources available to a protection domain (e.g., to eliminate ambient authority~\cite{miller2003capability} and the confused deputy problem~\cite{hardy1988confused}), thereby enforcing PoLP and serving as a prerequisite for formal verification to achieve high trustworthiness.
Consequently, modern microkernels such as seL4~\cite{heiser2020sel4}, L4Re~\cite{l4re}, and HongMeng~\cite{hm2024} all adopt capability systems as a core mechanism.
Unfortunately, conventional microkernel capability systems rely on hardware-based isolation to guarantee capability unforgeability, which incurs significant performance overhead when accessing capability-protected resources.
Accessing a capability-protected kernel object in seL4 consumes 379 CPU cycles (\autoref{tab:counters}), and this overhead primarily consists of CPU mode switching, capability lookup within the capability container (CNode),  capability decoding (necessitated by the semantic gap between the user-space application and the kernel regarding capability metadata), and authority check.

\begin{table}[htbp]
\setlength{\abovecaptionskip}{1.5pt} 
\setlength{\belowcaptionskip}{-2pt} 
\centering
\caption{Profile of the access of a kernel object in seL4 on the Raspberry Pi 3B+ board (in CPU cycles).}
\label{tab:counters}
\small
\resizebox{\columnwidth}{!}{
\begin{threeparttable}
\begin{tabular}{ccccc}
\toprule
\textbf{Mode switch} & \textbf{Cap lookup} & \textbf{Cap decode \& authority check} & \textbf{Others\tnote{\(\dagger\)}} & \textbf{Total} \\
\midrule
139 & 58 & 66 & 116 & 379 \\
\bottomrule
\end{tabular}
\begin{tablenotes}[flushleft]
  \footnotesize
  \item[\(\dagger\)] {Remaining overhead (e.g., IPC-buffer lookup, extra capability lookup, and return-path scheduling).}
\end{tablenotes}
\end{threeparttable}
} 
\end{table}

\begin{table}[htbp]
\setlength{\abovecaptionskip}{1.5pt} 
\setlength{\belowcaptionskip}{-7pt} 
\centering
\caption{Comparison of safe language-based kernels in terms of minimality principle adherence and TCB size.}
\label{tab:comparison}
\small
\resizebox{0.99\columnwidth}{!}{
\begin{threeparttable}
\begin{tabular}{
    m{2.2cm}
    >{\arraybackslash}m{3.7cm}
    >{\raggedleft\arraybackslash}m{1.8cm}
    >{\raggedleft\arraybackslash}m{0.8cm}
}
\toprule
\multirow{2}{2.2cm}{\centering\textbf{OS}} &
\multirow{2}{3.7cm}{\centering\textbf{Minimality}} & 
\multicolumn{2}{c}{\textbf{LCS in the TCB}} \\
\cmidrule(l){3-4}
& & \textbf{\texttt{core} + \texttt{alloc}} & \textbf{Others\tnote{\(\dagger\)}} \\
\midrule
\textbf{SPIN~\cite{spin1995}} & \ding{55} (memory mgr, thread mgr) & - & - \\
\textbf{Singularity~\cite{singularity07}} & \ding{55} (memory mgr, process mgr,\newline \hspace*{1em} IO mgr, scheduler) & - & - \\
\textbf{Tock~\cite{tock17,tock25}} & \ding{55} (process mgr, scheduler) & 1 911 & 2 903 \\
\textbf{RedLeaf~\cite{redleaf20}} & \ding{55} (memory mgr, scheduler) & 4 357 & 17 240 \\
\textbf{Theseus~\cite{theseus20}} & \ding{55} (memory mgr, scheduler) & 8 720 & 40 459 \\
\textbf{Asterinas~\cite{peng2025asterinas}} & \ding{55} (phy mem mgr) & 9 442 & 11 090 \\
\bottomrule
\end{tabular}
\begin{tablenotes}[flushleft]
  \footnotesize
  \item[\(\dagger\)] Linked code other than Rust \texttt{core} and \texttt{alloc}, including kernel/framework code and its direct or indirect unsafe dependencies required by the whole OS.
\end{tablenotes}
\end{threeparttable}
} 
\end{table}

The second issue is that the TCB of such kernels tends to become large.
First, beyond the kernel code, the TCB includes toolchain libraries and third-party dependencies linked into the kernel image.
Second, these kernels typically require all unsafe code to be encapsulated within the kernel, as allowing unsafe operations in protection domains would compromise the safe language-based isolation guarantees. 
We use the \emph{linked code size} (LCS) metric~\cite{peng2025asterinas} to count the total \emph{lines of code} (LoC) linked into the kernel image.
As detailed in \autoref{tab:comparison}, our evaluation of the functionality and LCS of several popular Rust-based kernels reveals that they do not follow the minimality principle, resulting in relatively large TCBs. Although Tock exhibits a smaller LCS, this is because it is a lightweight OS for low-power microcontrollers.

To address the above issues, we propose safeL4, a Rust-based single-address-space microkernel. The core idea of safeL4 is to construct capabilities that do not rely on hardware isolation, treating them as a first-class abstraction to uniformly provide protection domains with system resource abstractions that satisfy the microkernel minimality principle and fine-grained access control requirements.
First, based on Rust's object capabilities (i.e., object references or pointers~\cite{dennis1966programming,boyland2001cap,haller2010cap}), safeL4 constructs a capability system that does not require address space isolation for protection. It enforces fine-grained access control over system resources within a \emph{single address space} (SAS), thereby reducing the performance overhead caused by crossing address spaces in traditional capability systems.
Second, it adopts a microkernel approach to provide low-level abstractions for hardware resources and the communication functionality, and safely encapsulates only the essential unsafe operations, thereby reducing the TCB. 
Third, by utilizing a \emph{capability derivation tree} and Rust's visibility control, it solves the problem that Rust's native object capabilities can neither record capability derivation relationships nor be revoked.
This equips the capability system with interfaces of sufficiently rich semantics to support the kernel's management of system resources.

The contributions of this paper are summarized as follows.
\begin{itemize}
\item We propose an approach to constructing a capability system for system resources based on Rust's features, which avoids the heavy overhead of hardware isolation-based capability systems and satisfies microkernels' minimality principle as well as the requirements for capability-based fine-grained access control.

\item We address the issue that Rust's native object capabilities fail to meet the management requirements of microkernels for system resources.

\item 
We build a prototype of safeL4 in Rust and implement core OS services. We support applications written in unsafe languages to run in user space. 
Evaluation results show that safeL4’s safe abstraction achieves \(1.91\times\) the performance of seL4 on average, with IDC performance improving by \(7.8\times\) to \(22.3\times\).
\end{itemize}

%% file: sections/2.background.tex
\section{Background}

\subsection{seL4 Capability System}
The seL4 capability system governs access control for all kernel objects.
Any call on a kernel object must be made through a capability for that object, which encapsulates a reference to the object and a specific set of access rights that dictate the operations an application is authorized to perform.
By controlling the dissemination of these capabilities, the system enforces access control to kernel resources. The seL4 kernel uses a variant of the take-grant model~\cite{tg2008,tg2009} to manage the propagation of authority via capabilities. 
In seL4, capabilities are stored in the kernel to prevent tampering by user-level programs. 
The collection of capabilities owned by an application constitutes its capability address space (CSpace). Each capability is assigned a unique address (index) that the application uses to reference it.

\subsection{Object-Capability System}
An object-capability system~\cite{objcap2000,emily2006,objcap2017} is a security model that combines the principles of object-oriented programming with capability-based access control~\cite{levy2014capability} to manage and restrict access to resources within a software approach. 
In this model, objects are treated as resources, object methods serve as interfaces to access those resources, and object references act as capabilities. In Rust, for instance, references and smart pointers can fulfill this role.
Several memory-safe languages support object-capability models, including Java~\cite{javaobj10,javaobj24}, E~\cite{eobj06}, Encore~\cite{brandauer2015parallel}, Pony~\cite{clebsch2015deny}, and Rust~\cite{objcapcom2019}.

The unforgeability of capabilities is fundamental to preventing unauthorized access. 
Consequently, an object-ca\-pa\-bil\-i\-ty system must guarantee that object references cannot be forged.
This requirement explains why, when using Rust, only its safe subset can enforce an object-capability model: unsafe Rust allows raw pointer manipulation that could trivially bypass reference unforgeability. For this reason, safeL4 mandates that all code executing within its isolated domains be written exclusively in safe Rust.

%% file: sections/3.motivation.tex
\section{Related Work and Motivation}

\subsection{Safe Language-Based OSes} \label{sec-motiv-mks}

\begin{table*}[t]
  \caption{TCBs and safety guarantees of representative safe language-based operating systems.}
  \label{tab:safe-language-os-comparison}
  \centering
  \begingroup
  \footnotesize
  \setlength{\tabcolsep}{3pt}
  \renewcommand{\arraystretch}{1.08}
  \begin{tabularx}{\textwidth}{
    @{}
    >{\raggedright\arraybackslash}p{0.13\textwidth}
    >{\raggedright\arraybackslash}p{0.49\textwidth}
    >{\raggedright\arraybackslash}X
    @{}
  }
    \toprule
    \textbf{OS / language} &
    \textbf{Trusted Computing Base} &
    \textbf{Safety guarantees} \\
    \midrule

    \multicolumn{3}{@{}l}{\textit{Earlier safe language-based OSes (1990s - 2000s) }} \\
    \addlinespace[1pt]

    \textbf{SPIN}\newline Modula-3 &Modula-3 compiler and linker; Modula-3 runtime, including its garbage collector, and standard libraries; SPIN core OS services; DMA-capable devices and their drivers.&
    Isolation between protection domains (named safe extensions); capability protection on system resources. \\
    \addlinespace[2pt]

    \textbf{Singularity}\newline Sing\# &
    Bartok compiler and MSIL type-safety verifier; per-SIP runtime support, including its GC; Singularity microkernel code; DMA-capable devices and their drivers. & Isolation between protection domains (named SIPs); single ownership guarantee for exchanged objects; capability protection for system resources; contract-based communication between domains. \\
    \addlinespace[2pt]

    \textbf{House}\newline Haskell &
    GHC compiler and linker; modified GHC runtime, including its GC, and Haskell libraries; H monad (a hardware access interface layer), including its C and assembly code; DMA-capable devices and their drivers. 
    & Formally defined hardware access interfaces (excluding DMA devices) that enforce memory safety and support building OSes in functional languages (Haskell). \\
    \addlinespace[2pt]

    \textbf{J-Kernel}\newline Java &
    JVM bytecode verifier; JVM execution engine, GC, and Java system classes; J-Kernel library, including its class loader; host OS and devices. 
    & Isolation between protection domains; capability protection for exchanged objects. \\
    \addlinespace[4pt]
    \multicolumn{3}{@{}l}{\textit{Rust-based OSes (2015 - )}} \\
    \addlinespace[1pt]

\textbf{Tock}\newline Rust &
Rust compiler and linker; Rust \texttt{core} library; kernel code and its dependencies; DMA-capable devices and their drivers. 
& Hardware-based isolation between processes; isolation between protection domains (named capsules).\\
\addlinespace[2pt]

\textbf{Redleaf}\newline Rust &
    Rust compiler, linker and RedLeaf IDL compiler; Rust \texttt{core}/\texttt{alloc}, allocator; 
    kernel code base and its dependencies; DMA-capable devices and their drivers.
    & Isolation between protection domains; single ownership guarantee for exchanged objects. \\
    \addlinespace[2pt]

    \textbf{Theseus}\newline Rust &
    Rust compiler and linker; Rust \texttt{core}/\texttt{alloc}, allocator; kernel code base and its dependencies; DMA-capable devices and their drivers.
    & Isolation between protection domains (named cells); minimizing cross-domain state spill.  \\
    \addlinespace[2pt]

    \textbf{Asterinas}\newline Rust &
    Rust compiler and linker; Rust \texttt{core}/\texttt{alloc}, allocator; OSTD and its dependencies; IOMMU and devices outside its protection. 
    & Minimizing unsafe code and guaranteeing OS services are implemented in safe code and free of undefined behaviors. \\
    \addlinespace[2pt]

    \textbf{Atmosphere}\newline Rust&
    Verus frontend, Z3, Rust compiler and linker; Rust \texttt{core}, its formal specifications, and trusted axioms; proof helpers and
low-level system code; IOMMU and devices outside its protection. 
    & Machine-checked proofs establish functional correctness, memory safety, and leak freedom for the verified microkernel. \\
    \addlinespace[2pt]

    \bottomrule
  \end{tabularx}
  \endgroup

  \parbox{\textwidth}{%
    \footnotesize
    Entries of the \textit{Trusted Computing Base} column follow the same order: language or verification toolchain; language-specific runtime and libraries; trusted system code; and the IOMMU (or all devices when no IOMMU is used). All systems additionally assume correct CPU, memory, firmware, and boot behavior.}

  \Description{Comparison of representative safe-language operating systems by
  implementation language, main trusted components, and key system-level
  assurances.}
\end{table*}

Hardware-based isolation leverages the MMU to place each domain requiring isolation into a separate address space. In contrast, safe language-based isolation does not require a separate address space for each domain; instead, it runs all domains within a single address space, and domains can only communicate through their exposed interfaces.

Language-based isolation enables OSes to achieve better performance. It avoids the privilege-level switches and address-space switches introduced by hardware isolation, converting cross-domain communication into function calls. However, this approach changes the trust boundaries of OSes and the assumptions upon which their safety guarantees rely.
Hardware-based isolation relies on the CPU's MMU mechanism; therefore, OSes utilizing this type of isolation must assume the security of the CPU hardware. Conversely, the isolation provided by safe languages relies on the language's type system. Consequently, all components responsible for enforcing the type system must be included in the assumptions, including the compiler, standard libraries, and, for some languages, the runtime. As a result, the safe language-based isolation approach significantly increases the OS's TCB. We list the TCBs of representative safe language-based OSes and the safety guarantees they provide in~\autoref{tab:safe-language-os-comparison}.

Constructing kernels in safe languages has a long research history that can be divided into two phases. The first phase, spanning the 1990s and 2000s, explored using garbage-collected safe languages to build OS protection domains~\cite{bershad1994spin,spin1995,jkernel99,kaffeos00,golm2002java,hunt2005overview,fahndrich2006language,singularity07,hallgren2005principled}. 
Representative microkernels in this phase include SPIN~\cite{bershad1994spin,spin1995}, based on the Modula-3 language~\cite{modula3}, and Singularity~\cite{hunt2005overview,fahndrich2006language,singularity07}, based on Sing\#. 
Beyond providing protection domains, SPIN and Singularity also use the object capabilities of safe languages to design unforgeable tokens for system resources, serving as proofs of authority for resource access.
In addition, Singularity provides fault isolation for protection domains by enforcing that any memory shared between domains can only belong to one domain at a time (single ownership), preventing a crashing domain from affecting the rest of the system.

Contemporaneous with Singularity, there were also several OSes based on the Java language. J-Kernel~\cite{jkernel99} leverages Java’s type safety to establish boundaries between domains and designs a revocable capability for objects shared across domains, ensuring that any access from other domains to a crashed domain’s shared objects triggers an exception. KaffeOS~\cite{kaffeos00} implements a process abstraction within a \emph{Java virtual machine} (JVM) and provides isolated processes with mechanisms to manage resources such as CPU and memory. JX~\cite{golm2002java} implements a microkernel for managing hardware resources and uses the JVM as the unit of protection domains, running applications and services in different domains to achieve separation of privilege. However, because these safe languages rely on garbage collection, which can introduce unpredictable latencies, they conflict with the real-time requirements of some OS services~\cite{lee2013preemptible,emmerich2019case}.

The second phase began with Rust, which uses ownership, borrowing, and lifetimes to provide memory safety without garbage collection. This advantage introduces new development costs. Rust compiler's conservative borrow checker may reject safe programming patterns or require complex shared data structures to be reorganized, increasing development effort.
Moreover, avoiding a garbage collector does not eliminate unsafe code: Rust's \texttt{unsafe} mechanism enables operations beyond compiler checking, but makes programmers responsible for preserving Rust's safety invariants. Consequently, Rust-based OSes~\cite{tock17,tock25,redleaf20,theseus20,peng2024framekernel,peng2025asterinas,chen2025atmosphere} must carefully manage the distribution of unsafe code, because all such code lies within the TCB.
The first version of Tock~\cite{tock17} aimed to provide isolation for low-end embedded applications, and now it provides capabilities to restrict access to privileged APIs~\cite{tock25}. RedLeaf~\cite{redleaf20} captures inter-domain faults to achieve fault isolation. Theseus~\cite{theseus20} eliminates state dependencies~\cite{statespill} across domains. Asterinas~\cite{peng2024framekernel,peng2025asterinas} introduces a framekernel that encapsulates the unsafe Rust required for kernel development, thereby minimizing the TCB.



\subsection{Heavy Overhead of Current Microkernel Capability Systems}
Conventional microkernel capability systems rely on hardware isolation to ensure capability unforgeability, which imposes a significant performance overhead on microkernels. First, capability operations must be delegated to the kernel for execution via system calls, thereby introducing mode-switching overhead. 
Second, in the multi-server scenarios of microkernels, if a capability operation involves multiple protection domains, it also entails the cost of address-space switching. 
To avoid these overheads, current safe language-based kernels do not adopt the traditional microkernel's explicit, capability-centric authority model.
Although this approach reduces the performance overhead associated with capabilities, it weakens the capability system's advantages in areas such as the PoLP.

Safe language-based single-address-space microkernels provide a new opportunity to re-examine the tradeoff between performance and security.
First, object capabilities in safe languages no longer rely on hardware isolation to guarantee their unforgeability~\cite{dennis1966programming,boyland2001cap,haller2010cap}, which establishes the prerequisite for constructing a capability system within a single address space.
Second, the SAS design eliminates the costly CPU mode and context switches involved in resource access.
Furthermore, since protection domains and the kernel reside in the same address space, capabilities are invoked through typed interfaces, eliminating the overhead of decoding raw bytes to determine the capability type and the method being called.
Therefore, safe language-based microkernels can achieve the security benefits of a capability system without incurring a high overhead.

\subsection{New Minimality Requirement for Safe Language-Based OSes}
Minimality is the core principle of microkernels. In safe language-based microkernels, however, minimality entails another requirement besides minimizing kernel functionality: unsafe code usage must be minimized. This is because the prerequisite for the isolation provided by a safe language is that the code within a protection domain must be implemented using only safe code. 
Otherwise, the type-safe guarantees of the language could be compromised. Consequently, any component containing unsafe code inherently becomes the system's TCB.
Therefore, safe language-based microkernels must not only retain minimal hardware resource abstractions, but also design these abstractions as safe abstractions with sufficiently rich semantics. 
This enables protection domains residing in the same address space, particularly OS services such as file systems and device drivers, to be implemented entirely in safe Rust.

Fortunately, this presents a natural opportunity.
The hardware resources that a microkernel must manage are precisely where unsafe code is most concentrated in a Rust OS.
Therefore, these two requirements can be unified by encapsulating hardware resources as microkernel-style kernel objects, centralizing them within a minimal kernel that contains the unsafe code, and providing them to protection domains via safe interfaces.
Furthermore, modern microkernels, such as seL4, demonstrate that high expressiveness can coexist with strict minimality, enabling the construction of many practical OSes based on a small TCB (on the order of 10k LoC)~\cite{sel4cantripos,Kry10,magnetite,gapfruitos,sel4microkit,camkesvm,lionsos,sel4nio}.

Asterinas has already begun exploring the separation of safe and unsafe code within the intra-kernel address space.
By encapsulating unsafe hardware operations within a privileged framework, it constructs the kernel's memory-safety TCB in a principled manner. Asterinas aims to guarantee the memory-safety property of the entire kernel in a minimal way, thereby answering which unsafe operations must remain within the trusted framework. 
However, it still retains certain OS functionalities within the TCB, such as the physical memory manager, and it does not use capabilities for the unified granting, derivation, and revocation of authority over all system resources. 
Consequently, its privileged framework resembles an expressive safe OS development kit more than a microkernel constrained by principles such as minimality and PoLP.

%% file: sections/4.design.tex
\section{Design}

\subsection{Overview}
\begin{figure*}[htbp]
\setlength{\abovecaptionskip}{3pt}
\centering
\includegraphics[width=1.0\textwidth]{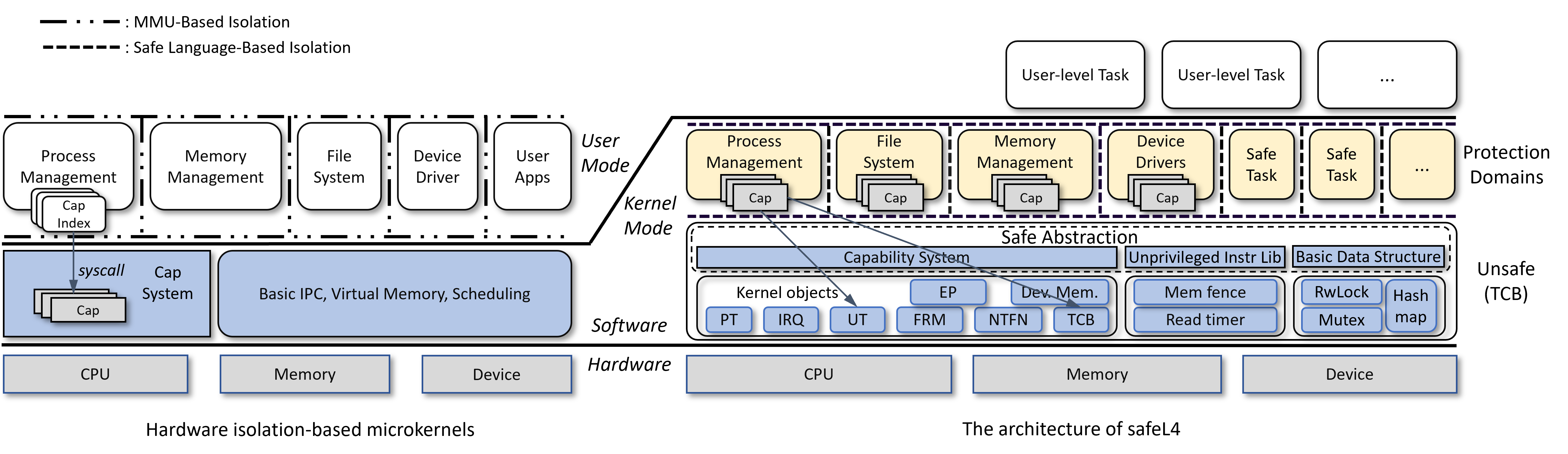}
\caption{Comparison between hardware isolation-based microkernels and safeL4 (EP: \cmttfont{Endpoint}; Dev.\ Mem.: \cmttfont{DeviceMemory}; PT: \cmttfont{PageTable}; IRQ: \cmttfont{Interrupt}; UT: \cmttfont{Untyped}; FRM: \cmttfont{Frame}; NTFN: \cmttfont{Notification}).}
\label{fig:sys_arch}
\end{figure*}

safeL4 is a safe language-based microkernel (see the right side of \autoref{fig:sys_arch}). The safeL4 microkernel abstracts system resources as kernel objects and encapsulates them with capabilities. A protection domain uses a Rust crate (a Rust compilation unit) as an isolation boundary. 
All protection domains run in the same address space and at the same privilege level (kernel mode) as the microkernel.
safeL4 requires that protection domains be implemented only in safe Rust, and that their use of hardware resources and IDC through the capabilities.
safeL4 supports two types of applications: safe tasks and user-level tasks. 
Safe tasks are OS services or applications that are written in safe Rust and run within protection domains. 
User-level tasks run in user space and are isolated by hardware, supporting legacy applications written in unsafe languages that invoke OS services within protection domains via POSIX system calls.

safeL4 builds a capability system based on Rust smart pointers, enforcing fine-grained access control over all system resources. safeL4 adheres to the minimality principle of microkernels by adopting a microkernel-like approach that provides low-level safe abstractions of hardware resources and by providing safe abstractions for non-privileged instructions and the basic data structures that must be implemented with unsafe Rust. These abstractions ensure that protection domains can implement their logic entirely in safe Rust. Furthermore, safeL4 confines a domain’s failures within itself through fault isolation.

\subsection{Trusted Computing Base and Safety Guarantees}\label{sub:TCB}

The safety of a safeL4-based OS relies on a trusted computing base and a set of execution assumptions. Under these assumptions,
safeL4 provides protection domains with two safety guarantees: language-level memory and type safety, and capability safety.

\textbf{Trusted components and assumptions (TCB).}
(1) We trust the Rust compiler to enforce the language-safety
guarantees of safe Rust, and the linker to correctly produce the final binary.
(2) We trust Rust's \texttt{core}/\texttt{alloc} libraries, the
global heap allocator, and the safeL4 microkernel and its
dependencies to correctly implement their safe interfaces.
(3) We assume that every protection domain and all of its
third-party dependencies are implemented in safe Rust.
The only trusted dependencies permitted to contain unsafe code are
Rust's \texttt{core}/\texttt{alloc} libraries and the safeL4
microkernel and its dependencies. 
(4) We trust the boot firmware, loader, and safeL4's initialization code to establish a well-formed initial state. 
(5) We assume that the processor, memory, platform hardware, and
peripheral devices behave according to their specifications.
Because unrestricted DMA can bypass Rust's language safety, safeL4 trusts DMA-capable devices, their drivers, and DMA configurations to confine DMA to authorized buffers, consistent with seL4's memory-access and DMA assumptions. The current safeL4 does not support IOMMU-based DMA isolation. A future extension of safeL4 to isolate devices via a correctly configured IOMMU would remove DMA-capable devices and their drivers from the TCB.

\textbf{About four \textit{safe} notions related to this paper.} 
\ding{182}  \textit{General safety} refers broadly to safety properties such as functional correctness and crash freedom.
\ding{183}  \textit{Safe Rust} refers to Rust code that itself does not contain \texttt{unsafe} code and benefits from the safety guarantees provided by the Rust language, such as memory safety, type safety, and data-race freedom.
\ding{184}  \textit{Safe Rust interfaces} (e.g., Asterinas's OSTD interfaces and \textit{safe abstractions} presented in this paper) have safe function signatures, but their implementations may contain unsafe code. Nevertheless, their implementations must semantically satisfy Rust’s safety requirements~\cite{rustbelt2017}.
\ding{185} \textit{Safe Rust over device interfaces} refers to safe Rust code that controls devices through \textit{safe Rust interfaces}. 
Although the code itself is written in safe Rust, hardware-induced changes to device state may undermine Rust's safety guarantees (e.g., the MMIO access use case in~\ref{sec:safety-case-studies}).


\textbf{Safety guarantees.} safeL4 provides the following two safety guarantees to protection domains.

\textit{Language-level memory and type safety.}
safeL4 encapsulates its \texttt{unsafe} code behind safe abstractions (safe notion~\ding{184}), allowing protection domains to be written in
safe code and thereby retain Rust language's safety (safe notion~\ding{183}). 
For safe Rust over MMIO devices, safeL4 provides the safe abstraction \cmttfont{DeviceMemory}, but it does not guarantee that drivers program devices correctly and such \textit{non-guarantees} are discussed below.

\emph{Capability safety.}
Capability safety mainly includes two aspects. 
(1) \emph{Capability-protected resource access.} Capabilities act as tokens to kernel objects that abstract hardware and communication resources. Rather than exposing raw pointers or hardware directly to a protection domain, a capability authorizes method invocations on its corresponding object, strictly governed by its associated rights.
(2) \emph{Safety properties of the capability system.}
As detailed in \autoref{sec-design-cap-security}, safeL4 enforces capability
unforgeability and non-bypassability, tracks capability provenance, preserves rights monotonicity, and ensures revocation correctness. These properties govern all invocations on capability-protected kernel objects. 

\textbf{Non-guarantees.}
safeL4 does not guarantee general safety (safe notion~\ding{182}) for protection domains, such as functional correctness. It also does not guarantee that driver code written in safe Rust (safe notion~\ding{185}) programs devices correctly: such code may use the \cmttfont{DeviceMemory} provided by the safeL4 microkernel to corrupt memory or compromise the system, for example through DMA misuse or violations of device protocols. We note that higher-level device-safety guarantees, such as preventing device-protocol violations, could be layered on top of our \cmttfont{DeviceMemory} API, for example via typestate checks as in Abacus~\cite{potyondy2026rage}.

\subsection{Capability System}
The capability system encapsulates safe abstractions of system resources and serves as the interface that safeL4 exposes to protection domains.
\subsubsection{Capability System Design} \label{sec-design-cap}
In Rust, the combination of \texttt{Arc} and \texttt{Weak} smart pointers is a common pattern for shared ownership.
Its semantics fit the basic sharing requirements of capabilities, and many Rust-based operating systems use this native mechanism to implement capabilities for shared resources. 
However, for a strict capability system, such as one that supports the take-grant model~\cite{harrison1976protection,tg1977}, directly using \texttt{Arc} and \texttt{Weak} as capabilities introduces the following two issues.

\textbf{Issue 1: \texttt{Arc}/\texttt{Weak} cannot support revocation semantics.}
In a capability system, subjects must be able to revoke resources they have shared.
However, the semantics of \texttt{Arc} prevent such revocation. Once subject A duplicates an \texttt{Arc} and passes it to subject B, A cannot perform a revocation as long as B still holds that \texttt{Arc}.
Additionally, the fact that \texttt{Arc}/\texttt{Weak} cannot represent multi-level capability derivation hierarchies prevents revocation. 
For example, suppose domain A owns a kernel object through an \texttt{Arc} and shares it with domain B as a \texttt{Weak}, and domain B then shares it with domain C. \texttt{Arc}/\texttt{Weak} cannot track the derivation relationship between domains B and C. As a result, in the sharing chain ``A$\rightarrow$B$\rightarrow$C'', domain B cannot revoke domain C’s access to the kernel object.

\textbf{Issue 2: \texttt{Weak::upgrade} prevents revocation.} 
When a domain calls \texttt{Weak::upgrade()}, it obtains an \texttt{Arc} and can hold the corresponding kernel object indefinitely. 
While this behavior is acceptable in common Rust programs, it is problematic in a capability system: 
if a protection domain can hold the upgraded \texttt{Arc} indefinitely, the original resource owner cannot revoke it.

To address the two issues, we combine Rust's \texttt{Arc}/\texttt{Weak} with a \emph{capability derivation tree} (CDTree) and propose the capability system as follows (\autoref{lst:cap_def}).
\ding{182} Each kernel object is wrapped in an \texttt{Arc<RwLock<KObj>\/>}, called \texttt{RawCap}, which ensures synchronized concurrent access to the object.
\ding{183} The capability system does not expose \texttt{RawCap} directly to protection domains. Instead, it provides a wrapper called \texttt{Cap}. \texttt{Cap} is a public struct visible to protection domains, but its internal fields are marked \texttt{pub(crate)}, restricting their visibility only to the microkernel.
\ding{184} The microkernel maintains the CDTree, a private global tree composed of \texttt{CdtNode}, to record derivation relationships among capabilities.
For all capability operations (e.g., derive, mint, revoke), the capability system maintains the following invariant: each node (\texttt{CdtNode}) in the CDTree has a 1:1:1 correspondence with a \texttt{RawCap} and its associated \texttt{Cap}.
\ding{185} The microkernel exposes a procedural macro that allows protection domains to invoke methods of the kernel object by a capability.
\ding{186} Each \texttt{Cap} contains a private field \texttt{rights} that specifies its authorities. 

\begin{figure}[htbp]
\setlength{\abovecaptionskip}{3pt}
\centering
\includegraphics[width=1\columnwidth]{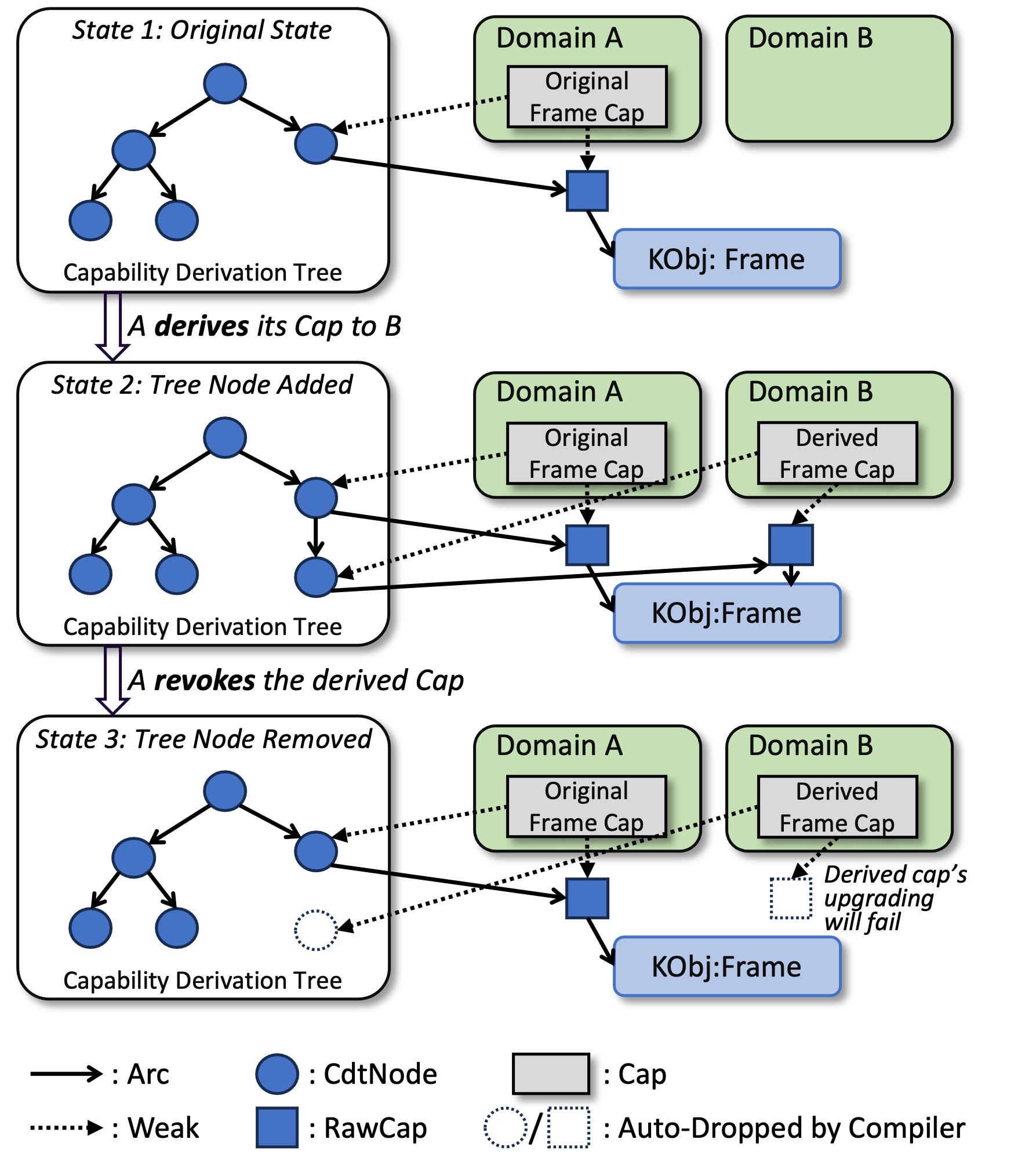}
\caption{Derivation and revocation of the capability system.}
\label{fig:capsys}
\end{figure}

For \textbf{Issue 1}, our solution is to enforce revocation semantics by maintaining the derivation relationships when resources are shared via \texttt{Arc}/\texttt{Weak}. 
We bind CDTree maintenance to the \texttt{Cap} operations built on \texttt{Arc}/\texttt{Weak}, ensure the CDTree retains ownership of the \texttt{RawCap}, and restrict domains to holding only a \texttt{Weak} to the \texttt{RawCap} through the \texttt{Cap}.
\autoref{fig:capsys} shows the changes in the CDTree when performing capability derivation and revocation operations.

\begin{lstlisting}[
    float=t,  % 作为单栏浮动体，优先放在栏顶
    caption={Struct definitions in safeL4's capability system.}, 
    captionpos=b, 
    aboveskip=3pt,
    belowskip=-2pt,
    abovecaptionskip=0pt,
    label={lst:cap_def}
]
pub(crate) struct RawCap {
    pub(crate) object: Arc<RwLock<KObj>>,
}
pub(crate) struct CdtNode {
    pub(crate) cap: Arc<RawCap>,
    pub(crate) child: Vec<Arc<Mutex<CdtNode>>>,
}
pub struct Cap {
    pub(crate) raw_cap: Weak<RawCap>,
    pub(crate) rights: Rights,
    pub(crate) cdt_node: Weak<Mutex<CdtNode>>,
}
\end{lstlisting}

\textit{State 1} $\rightarrow$ \textit{State 2}: When domain A derives a \texttt{Cap} for domain B, 
the capability system clones the \texttt{RawCap} and creates a new \texttt{CdtNode} to hold it. It then uses this \texttt{CdtNode} and the cloned \texttt{RawCap} to create a new \texttt{Cap}, returning this new \texttt{Cap} to domain A.
Domain A can then share the derived \texttt{Cap} with domain B.


\textit{State 2} $\rightarrow$ \textit{State 3}: When domain A revokes the \texttt{Cap}, 
the capability system locates the \texttt{CdtNode} via the \texttt{Cap} and drops the derived \texttt{CdtNode}. Consequently, the \texttt{RawCap} owned by the derived \texttt{CdtNode} is dropped as well. 
Thereafter, although domain B still holds the derived \texttt{Cap}, when it attempts to invoke the kernel object through this \texttt{Cap}, upgrading the \texttt{Weak} to the \texttt{RawCap} will fail.

For \textbf{Issue 2}, our key insight is that the capability system must limit the scope of \texttt{Weak} upgrades, and this enforcement must be guaranteed by the microkernel. 
safeL4 enforces this by making the \texttt{RawCap} visible only within the microkernel crate using \texttt{pub(crate)}, preventing protection domains from directly upgrading the \texttt{Weak} reference inside a \texttt{Cap}. 
safeL4 provides a procedural macro interface that redirects the kernel object method invocation to a public proxy function, which upgrades a \texttt{Cap}'s inner \texttt{Weak}~(see \autoref{sec-impl-capinvoke} for details). 
This ensures that all \texttt{Weak} upgrades occur only within the proxy function; thus, the lifetime of the upgraded \texttt{Arc}, and consequently the duration of resource access, is strictly controlled by the microkernel.

\subsubsection{Security Analysis of the Capability System} \label{sec-design-cap-security}
\mbox{}\\
\textbf{Capability unforgeability.}
In safeL4, protection domains are restricted to safe Rust, meaning they can only create capabilities through the capability system's public APIs. While safe Rust allows creating a raw pointer and performing pointer arithmetic (e.g., using \texttt{ptr::wrapping\_add}) to point to a capability or other sensitive data, an attacker cannot exploit these operations to break the safety enforced by safe Rust.
Any attempt to dereference such a raw pointer would require \texttt{unsafe} code blocks, which are disallowed in protection domains. Moreover, all safe interfaces of the safeL4 microkernel do not accept raw pointers as parameters. Therefore, attackers cannot convert a forged raw pointer into a valid capability through safeL4's safe interfaces.

\textbf{Protection domains cannot bypass the capability system.}
The pointer to the kernel object within a capability is encapsulated as a private field, preventing direct dereferencing in protection domains. 
In addition, each kernel object is defined as a private structure within the microkernel, ensuring that protection domains cannot instantiate or directly access kernel objects, nor can they directly call their methods to bypass the authority check. Consequently, the procedural macro provided by the microkernel is the only interface through which protection domains can invoke operations on the kernel objects referenced by capabilities.

\textbf{Provenance of a capability.}
Every capability is either created by the safeL4 microkernel for
a resource allocated during boot, or derived at runtime from an existing capability. For capability operations such as \texttt{copy}
and \texttt{revoke}, the capability system maintains a one-to-one
correspondence between each capability and a node in the CDTree.
Therefore, any capability can be traced along the CDTree derivation relation back to a root capability created during boot.

\textbf{Monotonicity of capability rights.}
All capability transition operations are implemented within the
microkernel. For derivation operations such as \texttt{copy} and
\texttt{mint}, the microkernel ensures that the rights of a derived
capability are a subset of those of its parent. Capability rights are
therefore monotonically non-increasing along every derivation path in
the CDTree.

\subsection{The Minimality Design}\label{subsub:min}
In the context of safe language-based microkernels, the minimality principle requires not only that the microkernel provide the minimum functionality, but also that it provide safe abstractions for all necessary unsafe operations, thereby ensuring that protection domains can be implemented exclusively in safe Rust.
To maintain minimality, safeL4 provides low-level safe abstractions for the following main categories of unsafe operations.

\textbf{Low-level safe abstraction for system resources.}
The first design principle for maintaining kernel minimality in safeL4 is to adopt the microkernel approach to constructing low-level abstractions for system resources. Inspired by seL4, we abstract fundamental hardware resources (CPU, memory, and I/O), privileged instructions, and communication functionality into corresponding kernel objects, and expose their interfaces as object methods. safeL4 requires the method signatures of kernel objects to be strictly safe, while their implementations may use unsafe code.

For CPUs, the \cmttfont{TCB} (\emph{thread control block}) object encapsulates access to and modification of CPU registers. 
Physical memory is abstracted as \cmttfont{Untyped} and \cmttfont{Frame} objects: \cmttfont{Untyped} objects represent allocatable physical-memory regions, while \cmttfont{Frame} objects provide methods for accessing and mapping frames. \emph{Memory-mapped I/O} (MMIO) devices are represented as \cmttfont{DeviceMemory} objects that provide safe methods for volatile read and write operations on device registers. 
Interrupts are represented by \cmttfont{Interrupt} objects, which can be bound to \cmttfont{Notification} objects for delivering interrupt events.
For privileged instructions, safeL4 follows seL4’s approach by wrapping them in the methods of kernel objects. OS services invoke these privileged instructions through these methods. 
We design the \cmttfont{Endpoint} object to handle synchronous communication, and use \cmttfont{Notification} objects for asynchronous communication. 
The low-level implementations of several of these objects use unsafe Rust because they manipulate raw memory or registers that Rust cannot validate. 
These safe abstractions are trusted to ensure that unsafe operations do not violate Rust's type safety guarantees, exposing safe interfaces to protection domains.

\textbf{Safe abstraction for non-privileged instructions.} 
The Rust \texttt{core} library provides wrappers for a subset of non-privileged instructions. 
For non-privileged instructions not covered by \texttt{core} that are required in device driver development, such as timer read and memory barrier instructions, safeL4 encapsulates them as safe functions in a library crate for protection domains. These wrappers use unsafe Rust because inline assembly may rely on conditions that rustc cannot fully verify. They are trusted to satisfy Rust's inline-assembly contract~\cite{rust_inline_assembly_rules} for every call.

\textbf{Safe abstraction for basic data structures.} Some fundamental data structures commonly used in Rust programs rely on unsafe code in their efficient implementations, including synchronization primitives (e.g., \texttt{Mutex}, \texttt{RwLock}) and heap-allocated collections (e.g., \texttt{HashMap}, \texttt{LinkedList}). The microkernel encapsulates these basic data structures and provides safe interfaces to protection domains. Rust's ownership rules make efficient implementations of shared mutable or pointer-linked data structures difficult using only safe Rust. Their unsafe implementations are therefore trusted to preserve the ownership, initialization, aliasing, and synchronization invariants promised by their safe interfaces.

Some unsafe code is not exposed to protection domains through safe abstractions, but is used internally to implement the safeL4 microkernel's own functionality. Typical examples include (1) the stack unwinder used in panic handling; (2) small amounts of assembly used in trap entry and return; and (3) unsafe code in the dependency tree of safeL4's procedural macros, which generates IDC and capability-invocation code.

Notably, safe abstractions do not eliminate unsafe code but rather confine it, providing safe interfaces to keep protection domains free of unsafe code. Therefore, any undefined behavior or arbitrary memory corruption introduced by unsafe code within them will still violate safeL4's safety guarantees.

\subsection{Fault Isolation}


Because the safe language-based microkernel and its protection domains share a single address space, a panic in one domain will propagate and crash the entire OS. 
Therefore, safeL4 must provide a fault isolation mechanism to capture panics permitted by the safe language, such as Rust's panics.

Inspired by some Rust-based OSes~\cite{theseus20,redleaf20,peng2025asterinas}, we implement a lightweight fault isolation mechanism.
When a protection domain invokes another domain via a synchronous IDC, safeL4 inserts fault isolation logic at the invocation point using a Rust procedural macro (see \autoref{sec-impl-idc} for details).
If a thread panics within the callee domain, the failure is captured at the call site. safeL4 unwinds the stack frames in the callee domain and cleans up the associated resources. The caller domain then receives an error from the IDC. 



%% file: sections/5.implementation.tex
\section{Implementation}

\subsection{The safeL4 Microkernel}
The safeL4 microkernel provides system resource operations to protection domains by implementing a set of kernel objects, including \cmttfont{TCB}, \cmttfont{Untyped}, \cmttfont{DeviceMemory}, \cmttfont{Frame}, \cmttfont{PageTable}, \cmttfont{MPI}, \cmttfont{Notification}, \cmttfont{Interrupt}, and \cmttfont{Endpoint}. 
All these kernel objects are accessed via capabilities. 
In addition, the microkernel provides safe interfaces for non-privileged instructions and basic data structures. 

\subsection{Inter-Domain Communication} \label{sec-impl-idc}
safeL4 supports both asynchronous and synchronous IDC. The asynchronous IDC is built on the \cmttfont{Notification} object. In addition to supporting the asynchronous semantics of seL4’s \cmttfont{Notification} (e.g., signal, wait, and poll), safeL4 extends it with a message-passing mechanism based on zero-copy ownership transfer, enabling efficient asynchronous communication between domains.

We implement synchronous IDC based on the \cmttfont{Endpoint} object. The callee domain uses link-time registration to register its interface with the microkernel, and then \cmttfont{Endpoint} uses the registered interface to invoke the callee domain.



\subsubsection{Link-Time Registration for Callee Crate}
Since the microkernel crate is a dependency of all protection domains, it cannot directly reference any domain crate. Otherwise, it would lead to a circular dependency.

To resolve this, we modify the linking phase of the microkernel and protection domains. Each callee crate uses \texttt{trusted\_kernel\_export!}, a Rust attribute macro provided by the microkernel, to annotate the functions it intends to export to other domains and the microkernel (\autoref{lst:IDCinterface}). 
During linking, the linker aggregates all exported interfaces from callee crates into a designated section of the compiled image. Each interface can be identified by a unique key consisting of the callee’s crate name and its function signature (including function name, parameter types, and return type).
When loading the callee image, the microkernel parses this section to obtain the interfaces of the callee and stores their function pointers for later invocation.

\begin{lstlisting}[caption={Example code of registering an interface.}, captionpos=b, label={lst:IDCinterface}]
// in MemoryManager (callee crate)
#[trusted_kernel_export]
pub fn ut_alloc_frame() -> Result<Cap, UtError> {
    UT_MANAGER.lock().do_alloc::<Frame>(FRAME_SIZEBIT)
}
// in ProcessManager (caller crate)
fn alloc_frame_cap() -> Result<FrameCap, UtError> {
    trusted_kernel_invoke!(mm::ut_alloc_frame() -> Result<Cap, UtError>).unwrap()
}
\end{lstlisting}



\begin{lstlisting}[
float=t,
caption={Simplified safeL4 microkernel workflow for resolving and calling IDC interfaces.}, captionpos=b, label={lst:invoke_proxy}, aboveskip=-3pt, belowskip=-3pt]
// trusted_kernel_invoke! calls this function after the type-match check
pub fn invoke_proxy<T>(
    api_identifier: ExportedAPIIdentifier,
    arg: *const (),
    ret: *mut (),
) -> Option<T> {
    GLOBAL_API_RESERVOIR.read()
    // find callee interface according to identifier tuple
    .get(&api_identifier)  
    .and_then(|interface| unsafe {
        // wrap callee execution
        crate::catch_unwind(|| { 
            interface.transmute_then_invoke(arg, ret);
        })
        .ok()
        .and_then(|_| {
            let ret_place_holder = &mut *(ret as *mut Option<MaybeUninit<T>>);
            ret_place_holder.take().map(|inner| inner.assume_init())
        })
    })
}
\end{lstlisting}

\subsubsection{Runtime IDC Invocation}
An \cmttfont{Endpoint} uses the procedural macro \texttt{trusted\allowbreak\_kernel\_invoke!} (\autoref{lst:invoke_proxy}) to invoke the callee domain’s interface.
The macro first performs a type matching to ensure that the parameters passed by the caller crate conform to the requirements of the callee interface. It then directly locates the corresponding callee interface using the compile-time generated unique key. Before transferring control to the callee domain, the invocation is wrapped in a \texttt{catch\_unwind} block to capture and handle potential panics in the callee domain. Additionally, safeL4 uses a custom compilation target to generate a section in the OS image for storing exception-handling information. Leveraging this information, safeL4 employs the \texttt{unwinding} library~\cite{Unwinding} to implement a stack-unwinding mechanism.


\subsection{Capability Invocation} \label{sec-impl-capinvoke}
A protection domain holds capabilities pointing to different kernel objects, and all accesses to a given kernel object must be performed by invoking the corresponding \texttt{Cap}. In safeL4, such capability invocation is exposed through \texttt{kobj!}, a procedural macro provided by the microkernel. The macro takes \texttt{Cap} as input and redirects the kernel object method to a public proxy function, which then performs the following steps:  (1) checks authority, (2) temporarily upgrades the \texttt{Weak} to the corresponding \texttt{RawCap}, and (3) tries to obtain the \texttt{RwLock} guarding the \texttt{KObj} inside the \texttt{RawCap}. The proxy function then executes the target method of the kernel object. After invocation, the proxy function releases the acquired \texttt{RwLock}, exits the scope of the \texttt{Weak::upgrade}, and returns.


\subsection{OS Services in Protection Domains}
We implemented fundamental OS services on safeL4, including process management, memory management, a file system, and several peripheral drivers.

\subsubsection{Process Management and Memory Management}
The \emph{process management} (PM) is responsible for creating, deleting, and scheduling all threads in the OS. PM uses the  \cmttfont{TCB} object to implement the thread abstraction for both safe tasks and user-level tasks, and implements a scheduler. In the multicore scheduling implementation, PM obtains the hardware core ID via the capability of an \cmttfont{MPI} object.


During system startup, the safeL4 microkernel initializes the physical memory reserved for protection domains as \cmttfont{Untyped} objects and transfers the corresponding capabilities to the memory management domain. The memory management domain implements a buddy system and a slab system based on these \cmttfont{Untyped} objects, and provides interfaces for memory allocation and deallocation to other domains.

\subsubsection{Porting RedoxFS}
RedoxFS~\cite{redoxfs} is a file system inspired by ZFS~\cite{bonwick2003zettabyte} that runs as a user-level process on Redox OS~\cite{redox}. We port RedoxFS and run it in a protection domain on safeL4, and the interfaces of RedoxFS are registered using \texttt{trusted\_kernel\_export!}. 

We remove all unsafe code blocks from the original RedoxFS code by leveraging our safe abstractions. RedoxFS performs in-place serialization and deserialization of inode data structures, which requires unsafe type conversions in Rust. To address this, inspired by PODType~\cite{pod}, we introduce a dedicated unsafe trait named \texttt{Convert} within the microkernel to support controlled type conversion and make it available for protection domains. We further enforce strict semantic constraints on this trait to restrict its behavior, which provides the following three additional guarantees.
\begin{enumerate}[label=(\arabic*)]
    \item Types that implement this trait cannot contain any function or data pointers, and all fields must be owned values.
    \item Types that implement this trait can only be converted to and from a byte array.
    \item The trait ensures that all type conversions occur in place with bound-checking to prevent out-of-bounds memory access during conversion.
\end{enumerate}

Since this trait is marked as \texttt{unsafe} within the microkernel, protection domains cannot implement it directly. Instead, protection domains use the \texttt{\#[derive(Convert)]} macro to mark their data structures that require the trait, and let the Rust compiler check if the requirements are satisfied and automatically derive implementations for these structures.


\subsubsection{Device Drivers}
The interrupt dispatch routine remains within the microkernel, while peripheral device drivers are implemented in protection domains. We have ported Redox’s SD card driver and implemented drivers for UART, timer, and DMA.
All device drivers access MMIO registers through the capabilities of their \cmttfont{DeviceMemory} objects, each of which represents a specific device memory region.

To support a new peripheral driver, we can create a \cmttfont{Device\allowbreak Memory} object pointing to the device’s MMIO region, pass the corresponding capability to the device driver’s protection domain during driver initialization, and bind the device’s interrupt (an \cmttfont{Interrupt} object) to the driver’s interrupt-handling IDC interface. 
For example, when porting Redox’s SD card driver, we replace its original unsafe operations, namely direct register access using \texttt{ptr::read\_volatile} and \texttt{ptr::write\_volatile}, with invocations on the capability of the \cmttfont{DeviceMemory} object (\autoref{lst:driver}).

\begin{lstlisting}[caption={safeL4 replaces unsafe MMIO access with invocations on the capability of the \cmttfont{DeviceMemory} object.}, captionpos=b, label={lst:driver}, belowskip=-3pt]
// Unsafe way to access MMIO area.
unsafe {(start as *mut T).add(offset).write_volatile(val);} 
// safeL4's cap-gated access (Offset is checked)
let result = kobj!(<device_memory_cap as DeviceMemoryCap>.write(offset, val)
    .map_err(|_| DMError::DMWriteError));
\end{lstlisting}

\subsubsection{Musl Libc Support}
To support user-level applications written in unsafe languages, we port musl libc to safeL4’s user layer and implement 50 system calls in OS services. A system call invoked by musl libc is transferred to the corresponding OS services.

%% file: sections/6.evaluation.tex
\section{Evaluation}
We evaluate safeL4's source code and linked code size, micro- and macrobenchmarks to achieve the following evaluation objectives, and use concrete examples from existing Rust OSes to evaluate safeL4's safety guarantees.

\begin{itemize}[leftmargin=*]
    \item \textbf{Objective 1.} By comparing safeL4 with seL4, evaluate the performance differences in low-level operations (such as IPC, interrupt handling, signal, and scheduling) between a safe language-based microkernel and a hardware isolation-based microkernel.

    \item \textbf{Objective 2.} Evaluate the performance differences between safeL4, seL4, and Linux when handling realistic workloads.

    \item \textbf{Objective 3.} Evaluate the file system-related microbenchmark performance between safeL4 and Redox, to reveal the impact of different microkernel architectures.
\end{itemize}

\subsection{Experiment Setup}

All experiments are conducted on a Raspberry Pi 3B+ board with four Cortex-A53 cores (1.4 GHz), 1 GB SDRAM, and a 32 GB SD card. We have set up the following OSes on this hardware platform to achieve our evaluation objectives.

\textbf{seL4.} seL4 is a hardware isolation-based microkernel with the fastest IPC. We evaluate the performance differences between safeL4 and seL4 in terms of their low-level abstraction of hardware resources (\emph{Objective 1}) and realistic workloads  (\emph{Objective 2}). The most common deployment scenario for seL4 is as a library OS (LibOS), in which resources are statically allocated, and most computations do not require system calls. We run YCSB benchmarks~\cite{ycsb} on SQLite3~\cite{sqlite} in the user space of safeL4 and seL4 LibOS.

\textbf{Redox.} 
Redox OS is a relatively mature microkernel-based general-purpose OS based on hardware isolation and implemented in Rust. We have ported RedoxFS and the SD card driver to safeL4's protection domains while maintaining source code consistency.
We run file system-related microbenchmarks in user space on both Redox and safeL4 to evaluate the performance difference between safe language-based isolation and hardware-based isolation (\emph{Objective 3}).

\textbf{Linux.} Linux is a representative monolithic OS. We use Ubuntu 22.04 (Linux 6.6.12) with the ZFS file system. We run SQLite3 in the user space of safeL4 and Linux (\emph{Objective 2}).

\subsection{Source Code and Linked Code Size Evaluation}

We evaluate both the source code size and the \emph{linked code size} (LCS) included in the final image. For the latter, we adopt the LCS metric used in Asterinas~\cite{peng2025asterinas}. However, it is important to note that the source code size and LCS do not fully represent the TCB of safeL4. A complete account must also incorporate the remaining assumptions specified in \autoref{sub:TCB}.

The entire safeL4 OS consists of 22,018 LoC (\autoref{tab:loc}). The microkernel contains 11,688 LoC, including 1,765 lines of unsafe code, which accounts for 15.1\% of the microkernel code. 
We support the RISC-V64 and AArch64 architectures.
\autoref{tab:loc} shows that the amount of unsafe code required to support a new CPU architecture is quite small, 462 LoC for AArch64 and 349 LoC for RISC-V64, respectively.


\begin{table}[htbp]
\setlength{\abovecaptionskip}{2pt}
\setlength{\belowbottomsep}{-8pt}
\centering
\caption{Lines of Code for safeL4.}
\label{tab:loc}
\small
\resizebox{\columnwidth}{!}{
\begin{tabular}{lS[table-format=5]|lS[table-format=5]lS[table-format=5]}
\toprule
&\textbf{LoC} & \textbf{Crate} & \textbf{LoC} & \textbf{Crate} & \textbf{LoC} \\
\hline
\textbf{Safe code}      & 9923  & Microkernel        & 11688 & DMA       & 328 \\
\textbf{Unsafe code: }  &       & Process management & 3943  & UART      & 261 \\
\quad Arch-independent  & 954   & Memory management  & 502   & RAM Disk  & 110 \\
\quad AArch64           & 462   & RedoxFS            & 4254  & Mailbox   & 81 \\
\quad RISC-V64          & 349   & SD card driver     & 777   & Timer     & 74 \\
\hline
\textbf{Total (Microkernel)} & 11688 & \textbf{Total} &      &           & 22018 \\
\bottomrule
\end{tabular}
}
\end{table}

We further evaluate safeL4's LCS. We count all source lines referenced in the linked IR generated by the LLVM toolchain. We evaluate the LCS of both the TCB components (primarily including the Rust \texttt{core} and \texttt{alloc} libraries, the safeL4 microkernel, and its direct and indirect crate dependencies) and the non-TCB components (protection domains).

The evaluation results (\autoref{tab:tcb}) show that (1) in Rust-based OSes, the Rust \texttt{core} and \texttt{alloc} libraries account for a large fraction of the trusted linked code. (2) Compared with other Rust-based OSes, safeL4’s low-level abstraction approach yields the smallest amount of other trusted linked code for system resource abstraction (others in \autoref{tab:tcb}). Although Tock also has a small amount of such code, it is a lightweight OS aimed at low-power microcontrollers.


\begin{table}[htbp]
\setlength{\abovecaptionskip}{2pt}
\centering
\caption{Comparison of LCS.}
\label{tab:tcb}
\small
\resizebox{0.9\columnwidth}{!}{
\begin{tabular}{lcrrr}
\toprule
\multirow{2}{*}{\quad \textbf{OS}}
& \multicolumn{2}{c}{\textbf{LCS in the TCB}}
& \multirow{2}{*}{\shortstack[c]{\textbf{LCS outside}\\[2pt]\textbf{the TCB}}}
& \multirow{2}{*}{\textbf{Total}} \\
\cmidrule(lr){2-3}
& \textbf{\texttt{core} + \texttt{alloc}} & \textbf{others} & & \\
\midrule
Asterinas & 9 442 & 11 090 & 58 165 & 78 697 \\
Theseus   & 8 720 & 40 459 &  9 595 & 58 774 \\
RedLeaf   & 4 357 & 17 240 &  4 813 & 26 410 \\
Tock      & 1 911 &  2 903 &  1 814 &  6 628 \\
safeL4    & 6 248 &  2 720 &  6 081 & 15 049 \\
\bottomrule
\end{tabular}
}
\end{table}

\subsection{Microbenchmark Comparison with seL4}
We use seL4bench~\cite{sel4bench} as a microbenchmark suite to compare the performance of safeL4 and seL4. seL4bench is designed to evaluate various low-level operations of the seL4 microkernel. On safeL4, we reimplement all seL4bench benchmarks in Rust, excluding VCPU (virtualization), and execute them in protection domains.

\begin{table*}[htbp]
\setlength{\abovecaptionskip}{2pt}
\centering
\caption{The seL4bench results in CPU cycles. Smaller is better.}
\label{tab:seL4bench}
\small
\resizebox{\textwidth}{!}{ 
\begin{tabular}{llS[table-format=5.3(6)]S[table-format=4.3(6)]S[table-format=4.2]}
\toprule
\textbf{seL4bench} & \textbf{Description} & \textbf{seL4} & \textbf{safeL4} & \textbf{Perf*} \\
\midrule
Fault           & hardware fault triggering-to-handling & 589.470 \pm 22.487 & 309.952 \pm 5.206 & 1.902 \\
Hardware        & calling into kernel and return & 149 \pm 34.795 & 44 \pm 0.000 & 3.386 \\
IPC/IDC         & invoke a server in a different address space (seL4)/domain (safeL4) & 249 \pm 0.000 & 32 \pm 0.000 & 7.781 \\
                & a server replies to a client in a different address space (seL4)/domain (safeL4) & 268 \pm 0.000 & 12 \pm 0.000 & 22.330 \\
IRQ             & invoke an interrupt handler in a user-thread (seL4)/domain (safeL4) & 622.440 \pm 21.835 & 413.632 \pm 1.502 & 1.505 \\
Page mapping    & prepare one page table        & 1354.438 \pm 120.572 & 1153.889 \pm 151.577 & 1.174 \\   
                & map one page                  & 390.377 \pm 4.420    & 260.900 \pm 12.518   & 1.496 \\   
                & allocate one frame capability & 538.345 \pm 24.992   & 949.200 \pm 14.119   & 0.567 \\   
Signal          & asynchronous signal delivery  & 406.050 \pm 22.614   & 113.684 \pm 1.243    & 3.572 \\ 
                & synchronous signal delivery   & 560.830 \pm 22.979   & 449.309 \pm 8.072    & 1.248 \\ 
Scheduler       & yield a thread                & 382.31 \pm 19.130    & 400.056 \pm 0.631    & 0.956 \\ 
                & schedule between two threads  & 758.100 \pm 15.400   & 584.955 \pm 3.462    & 1.296 \\
Synchronization & broadcast a signal through notification        & 1370.860 \pm 26.950  & 892.400 \pm 23.589    & 1.536 \\
                & the last waiter of a waiting chain being woken & 71296.000 \pm 14.949 & 45868.900 \pm 208.884 & 1.554 \\
                & producer to consumer latency  & 1217.740 \pm 140.721 & 876.600 \pm 9.088    & 1.389 \\
                & consumer to producer latency  & 1194.160 \pm 28.889  & 860.800 \pm 3.513    & 1.387 \\
SMP             & intra-core IPC round-trip     & 6888.61 \pm 0.410    & 4877.98 \pm 2.460    & 1.411 \\
\midrule
\textbf{Geometric Mean}  & & & & 1.910 \\
\bottomrule
\\[-2mm]
\multicolumn{5}{l}{* Normalized performance, use seL4/safeL4 to indicate safeL4's performance advantage over seL4. Higher is better.}\\
\end{tabular}
}
\end{table*}

\begin{table*}[htbp]
\setlength{\abovecaptionskip}{2pt}
\centering
\caption{File system-related comparison with Redox.}
\label{tab:redox_comparison}
\renewcommand{\arraystretch}{0.8}
\small
\resizebox{0.8\textwidth}{!}{
\begin{tabular}{llcS[table-format=3.3(4)]S[table-format=3.3(4)]S[table-format=3.2]}
\toprule
\textbf{LMbench} & \textbf{Description} & \textbf{Unit} & \textbf{Redox} & \textbf{safeL4} & \textbf{Perf*} \\
\midrule
memory map         & map then unmap a block of memory  & $\mu s$ & 27.789 \pm 0.049  & 21.795 \pm 0.052  & 1.275 \\
page fault         & triggers page fault for each page & $\mu s$ & 150.988 \pm 0.129 & 94.336 \pm 0.157  & 1.601 \\
file access        & open, read then close a file      & MB/s    & 8.269 \pm 0.053   & 18.142 \pm 0.009  & 2.194 \\
                   & read a file                       & MB/s    & 25.199 \pm 0.139  & 55.618 \pm 0.007  & 2.207 \\
memory map         & map, read, then unmap a file      & MB/s    & 345.429 \pm 7.180 & 380.940 \pm 0.000 & 1.103 \\
                   & open, map, read, unmap, close     & MB/s    & 8.287 \pm 0.001   & 12.082 \pm 0.002  & 1.458 \\
file create/delete & create files                      & files/s & 32.685 \pm 1.960  & 40.894 \pm 0.986  & 1.251 \\
                   & delete files                      & files/s & 74.854 \pm 4.370  & 82.138 \pm 1.833  & 1.097 \\
\midrule
\textbf{Geometric Mean} & & & & & 1.470 \\
\bottomrule
\\[-2mm]
\multicolumn{5}{l}{* Normalized performance. For throughput, use safeL4/Redox; for latency, use Redox/safeL4.  Higher is better.}\\
\end{tabular}
}
\captionsetup{skip=4pt}
\end{table*}

The evaluation results (\autoref{tab:seL4bench}) show that safeL4 is \(1.91\times\) (geometric mean) faster on average than seL4, with IDC demonstrating the most significant performance advantage, being \(7.8\times\) to \(22.3\times\) faster than seL4.
The results may help answer the question about the performance comparison between safe language-based microkernels and those that rely on hardware isolation~\cite{diffmk08}: for microkernels, the overhead introduced by Rust code (including type safety checks and fault isolation mechanisms) is lower than the overhead introduced by hardware isolation (including CPU context switching, privilege-level switching, and page table switching).


Synchronization-related benchmarks (signal/schedule/sync) also benefit from safeL4. In safeL4, the primary overhead of a synchronization operation comes from accessing CPU contexts through capabilities, CPU context switching triggered by thread switches, and atomic operations within semaphores. They do not involve CPU mode switching or address space switching, thereby outperforming seL4.

\begin{table}[htbp]
\setlength{\abovecaptionskip}{2pt}
\centering
\caption{Profile of allocation of a frame.}
\label{tab:alloc_frame}
\small
\resizebox{1.0\columnwidth}{!}{ 
\begin{tabular}{lr}
\toprule
\textbf{Operation} & \textbf{Avg. Cycles} \\
\midrule
Acquire the RwLock on a capability to access Untyped kernel object & 99.28  \\
Retype a Frame kernel object from the Untyped kernel object        & 165.32 \\
Create a Frame capability using the Frame kernel object              & 656.11 \\
Insert the Frame capability's CdtNode to its parent                & 29.00  \\
\bottomrule
\end{tabular}
}
\end{table}

\begin{figure}[htbp]
\setlength{\abovecaptionskip}{3pt}
\centering
\includegraphics[width=0.7\columnwidth]{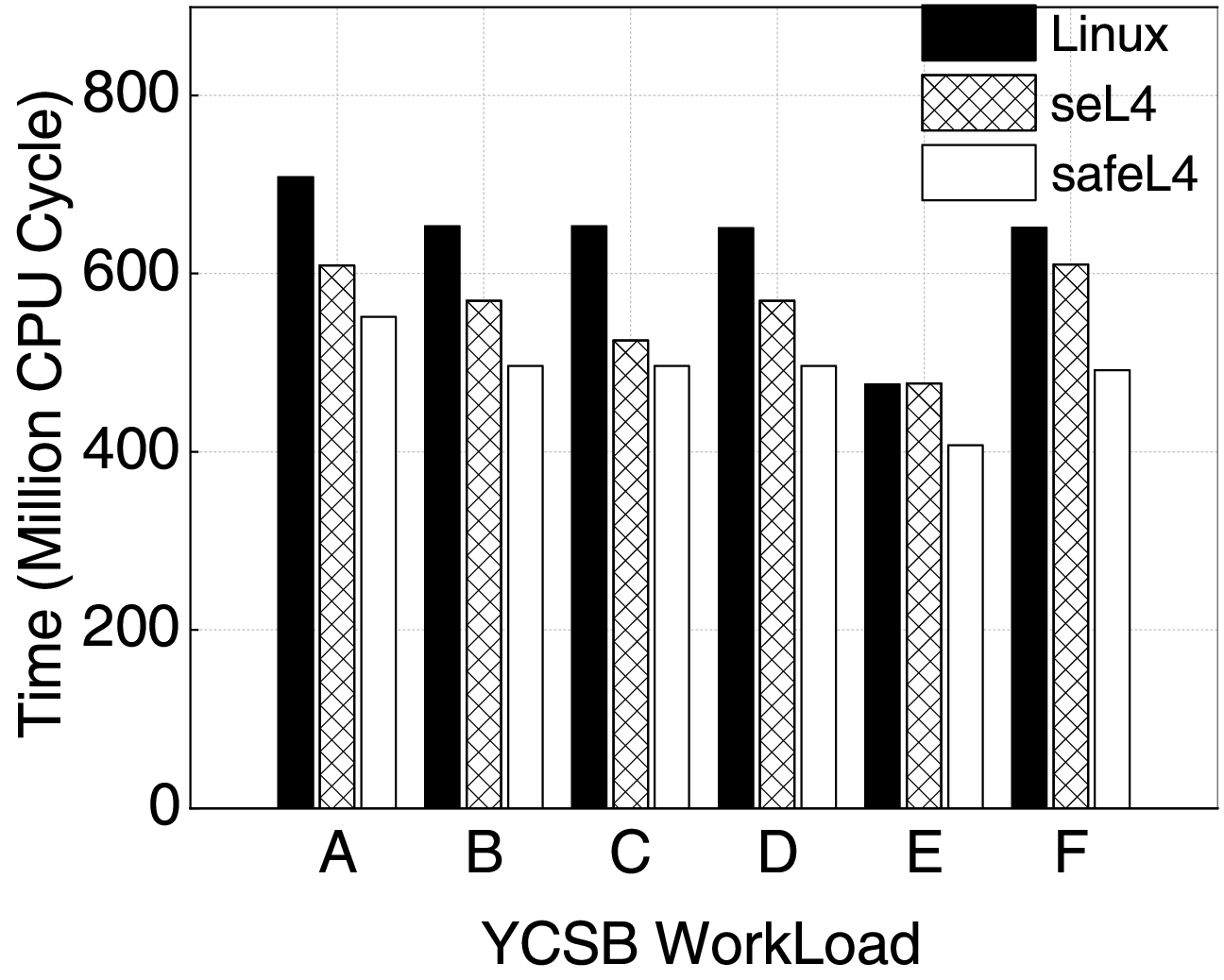}
\caption{SQLite3 YCSB benchmark, smaller is better.}
\label{fig:YCSB}
\end{figure}

Here, we explain why safeL4 underperforms seL4 in the frame allocation benchmark. This primarily stems from design divergences in concurrency strategies of kernel data structures. seL4 employs coarse-grained concurrency control via a big kernel lock, which ensures exclusive access, allowing kernel data structures to operate without internal synchronization checks. In contrast, safeL4 adopts a fine-grained locking design. Accessing kernel objects via capabilities in safeL4 requires acquiring a read-write lock (\texttt{RwLock}) to ensure correctness in concurrent environments, which inevitably introduces synchronization overhead at the granularity of kernel objects. We profile the synchronization overhead incurred by this fine-grained locking (\autoref{tab:alloc_frame}). The results indicate that during frame allocation, the initialization of a capability with \texttt{RwLock} and \texttt{Arc} consumes an average of 656.11 CPU cycles, constituting the dominant overhead. Furthermore, a portion of the frame allocation overhead in safeL4 originates from maintaining the CDTree.


\subsection{Macrobenchmark Comparison with seL4 and Linux}

We evaluate the performance of safeL4, seL4, and Linux using the YCSB benchmark on SQLite3. SQLite3 is ported to seL4 LibOS and deployed as a user-level task on safeL4. The workload comprises six standard YCSB scenarios, covering read, write, update, and mixed operations on a table containing 1,000 records.

The results (\autoref{fig:YCSB}) show that safeL4 outperforms seL4 by 14.3\% and Linux by 23.2\% on average. Both safeL4 and the seL4 LibOS achieve higher performance than Linux. 

To explain safeL4’s advantage over Linux, we profile system calls during the YCSB benchmark. Four system calls, \texttt{fcntl}, \texttt{newfstatat}, \texttt{pread64}, and \texttt{pwrite64}, account for more than 70\% of the total execution time. The system calls implemented by safeL4 are simpler than Linux’s. For example, there is much less locking logic (no byte-range locking and no need for a global lock linked list), fewer system states to maintain, no access control checks, and so on.


In this benchmark, the performance improvement of safeL4 over seL4 is not as good as in seL4bench, for the following reasons. First, SQLite3 and YCSB, as implemented in an unsafe language, run in user space on safeL4, so they need to allocate resources via system calls. Second, the seL4 LibOS adopts a static allocation strategy: the memory required by applications is allocated in one shot during initialization, reducing the number of system calls during execution.
Third, in the seL4 LibOS, SQLite3, the file system and drivers run in the same address space, and there is no IPC overhead between them. 
The second and third reasons also explain why seL4 outperforms Linux in this benchmark.

\subsection{Microbenchmark Comparison with Redox}
We port the RedoxFS and SD card driver from Redox OS, maintaining source and semantic consistency. In Redox OS, the file system and drivers run as user-level processes. In safeL4, they run in protection domains.

We cannot run LMbench directly on Redox because it only supports a limited set of POSIX system calls on AArch64. 
We implement an LMbench subset related to the file system in Rust and run it in user space on both safeL4 and Redox.

The results (\autoref{tab:redox_comparison}) show that safeL4 outperforms Redox in all file system-related microbenchmarks, with a geometric-mean speedup of \(1.47\times\). This experiment demonstrates that it is feasible to port existing typical OS services to safeL4 using its safe abstractions and that these OS services can benefit from safeL4’s single-address-space design.

\subsection{Case Study: MMIO Abstraction and Capability-Protected Device Access}
\label{sec:safety-case-studies}

We use two concrete device-access examples from existing Rust OSes to illustrate the benefits of safeL4's language-level memory/type safety, as well as capability safety.

\textbf{MMIO access without Rust references.}
Theseus's PL011 driver maps an MMIO region into a \texttt{MappedPages} object and borrows the mapping as a typed mutable register block.

\begin{lstlisting}[
  language=Rust,
  caption={Theseus constructs a typed mutable reference over a PL011
  MMIO mapping (simplified for brevity).},
  label={lst:theseus-mmio},
  captionpos=b,
  aboveskip=4pt,
  belowskip=0pt,
  abovecaptionskip=0pt,
]
// kernel/uart_pl011/src/lib.rs
// Pl011::new
let regs =
    mapped_pages.into_borrowed_mut::<Pl011_Regs>(0)?;
// kernel/memory/src/paging/mapper.rs
// MappedPages::as_type_mut
let typed: &mut T =
    unsafe { &mut *(start_vaddr as *mut T) };
\end{lstlisting}

This issue was previously identified in the Asterinas paper~\cite{peng2025asterinas}: Theseus's
\texttt{MappedPages} API eventually constructs an \texttt{\&mut T} over an MMIO mapping. This is unsound for externally modifiable memory because hardware may update device registers independently, whereas \texttt{\&mut T} assumes no external modifications.

safeL4 instead represents each MMIO region as a capability-protected \texttt{DeviceMemory} object. Its interface exposes neither Rust references nor raw pointers into the region. Each invocation causes the microkernel to perform a bounds-checked volatile read or write. Thus, under the assumptions in \autoref{sub:TCB}, MMIO access through this interface preserves language-level memory/type safety.

\textbf{Capability-protected device access.}
A sound MMIO interface alone does not determine which domain may access a particular device. RedLeaf, for example, relies on a PCI domain to discover PCI devices, map their memory-mapped I/O regions, and grant drivers access to those regions. However, any driver domain with access to this service can invoke \texttt{pci\_register\_driver}, claim another device's vendor and device identifiers, and request access to one of that device's MMIO regions.

\begin{lstlisting}[
float=t,
  language=Rust,
  caption={RedLeaf's PCI domain selects a device and BAR from caller
  inputs (simplified for brevity).},
  label={lst:redleaf-pci},
  captionpos=b,
  aboveskip=4pt,
  belowskip=0pt,
  abovecaptionskip=0pt,
]
// domains/sys/driver/pci/src/lib.rs
// PCI::pci_register_driver
let (vid, did) =
    (pci_driver.get_vid(), pci_driver.get_did());
let devices = PCI_DEVICES.lock();
let dev = devices.iter().find(|d|
    (d.vendor_id(), d.device_id()) == (vid, did)
)?;
let bar = dev.get_bar(
    bar_index, pci_driver.get_driver_type());
pci_driver.probe(bar);
\end{lstlisting}

The PCI domain thereby acts as a \textit{confused deputy}: a driver can exploit the PCI domain's broader authority to acquire and subsequently access a device-memory region that was not assigned to it. safeL4 instead directly grants the designated driver capabilities for a \texttt{DeviceMemory} object and the \texttt{Notification} associated with the device interrupt during initialization. The capability safety guarantees that the driver can invoke methods only on authorized objects for which it holds capabilities, removing the confused deputy issue.

\textbf{Guarantees and limitations.}
In safeL4, the safe abstraction over \texttt{DeviceMemory} constrains \emph{how} MMIO is accessed, while the capability system constrains \emph{which objects} a domain may invoke. However, these guarantees do not establish driver functional correctness: a driver may still contain logic errors and  violate device-specific state-machine protocols. 

%% file: sections/8.conclusion.tex
\section{Conclusion}
This paper presents safeL4, a safe language-based microkernel built around a capability-centered design. The key idea is to combine a single-address-space design for low-overhead capability invocation with safe language-enforced capability unforgeability, and to treat the capability as a first-class kernel abstraction for uniformly exposing system resources to protection domains. Our evaluation shows that safeL4’s primitives achieve 1.91$\times$ the performance of seL4 on average, and that although the Rust toolchain accounts for a notable portion of the TCB, the kernel itself remains compact.

%% file: sections/safeL4_ae_appendix.tex
%



\newcommand{\ArtifactURL}{\url{https://doi.org/10.5281/zenodo.22852236}}
\newcommand{\ArtifactSize}{Under 15 GB}
\newcommand{\EvaluationTime}{10-12 hours}
\newcommand{\HostLinuxVersion}{Ubuntu 22.04}
\newcommand{\SafeLFourImageLicense}{MIT}

\setlength{\parindent}{1em}
\setlength{\parskip}{0pt}

\appendix
\section{Artifact Appendix}

\subsection{Abstract}

This artifact provides a pre-built safeL4 image, together with the source
code and pre-built images for the seL4, Redox, and Linux. It supports the reproduction of three main experiments from the paper on a Raspberry Pi 3 Model B+: the microbenchmark
comparison with seL4, the SQLite3/YCSB macrobenchmark comparison with seL4 and
Linux, and the file-system microbenchmark comparison with Redox. Detailed
instructions for running each experiment are
included in the artifact repository.

\subsection{Artifact check-list}

{\small
\begin{itemize}
  \setlength{\itemsep}{1pt}
  \item \textbf{Program:} safeL4 image; seL4 and seL4bench; Redox OS;
        Ubuntu Linux with ZFS; YCSB benchmark on SQLite3; a file-system-related subset of LMbench.
  \item \textbf{Compilation:} Not required for the main evaluation; all
        evaluated systems are supplied as pre-built images. Source code
        of seL4, Redox, and Linux are included for inspection and optional
        rebuilding.
  \item \textbf{Run-time environment:} Bare-metal Raspberry Pi 3 Model B+;
        a separate Linux host is used to write images and capture UART output.
  \item \textbf{Hardware:} Raspberry Pi 3 Model B+, 32 GB or larger microSD
        card, card reader, and 3.3 V USB-to-TTL UART adapter.
  \item \textbf{Output:} UART logs and parsed numerical summaries.
  \item \textbf{Experiments:} Microbenchmark comparison with seL4;
        macrobenchmark comparison with seL4 and Linux; microbenchmark
        comparison with Redox.
  \item \textbf{Required disk space:} \ArtifactSize.
  \item \textbf{Experiment run time:} \EvaluationTime.
  \item \textbf{Public link:} \ArtifactURL
  \item \textbf{Licenses:} safeL4 image: \SafeLFourImageLicense;
        third-party sources and binaries retain their upstream licenses.
\end{itemize}
}

\subsection{Description}

\subsubsection{How to access}

The artifact and its detailed \texttt{README} are available at
\ArtifactURL. The artifact contains the safeL4 image; the source code and
images for seL4, Redox, and Linux; checksums; and documentation for building
and running the images, conducting the experiments, and extracting the
results.

\subsubsection{Hardware dependencies}

The evaluation requires:
\begin{itemize}
  \setlength{\itemsep}{1pt}
  \item one Raspberry Pi 3 Model B+ with four Cortex-A53 cores at 1.4 GHz and
        1 GB SDRAM, plus a suitable power supply;
  \item one 32 GB or larger microSD card and a USB card reader;
  \item one Linux host with an available USB port; and
  \item one 3.3 V USB-to-TTL UART adapter and jumper wires for the Raspberry Pi UART pins.
\end{itemize}

\subsubsection{Software dependencies}

The Linux host described above should run a 64-bit Debian-based Linux distribution (tested with \HostLinuxVersion) and have tools for checksum verification, writing raw images to microSD cards, capturing UART output, and parsing results. The required packages and installation commands are provided in the repository.

\subsection{Installation}

We provide pre-built images for all evaluated operating systems, so evaluators do not need to install the corresponding build environments. Instructions for running the OS images, as well as optional build instructions, are provided in the \texttt{README}.

\subsection{Experiment workflow}

To run an experiment, follow the instructions in the \texttt{README} in the corresponding experiment directory. First, flash the experiment-specific Raspberry Pi 3B+ firmware and OS images to the SD card. Insert the SD card into the Raspberry Pi, connect its UART interface to a host device that captures the serial output, and power on the board. Then follow the remaining instructions in the \texttt{README} to execute the experiment and collect the raw numerical results. Each experiment directory also provides scripts or documentation explaining how the raw results are processed or mapped to the  reported results.

\subsection{Evaluation and expected results}

Due to differences in hardware and execution environments, the exact results presented in the paper may differ from those reproduced on other machines. However, the relative performance trends and conclusions drawn in our evaluation should hold.

\subsubsection{Microbenchmark Comparison with seL4}

This experiment compares safeL4 with seL4 to assess the performance differences in low-level operations, including IPC, interrupt handling, signaling, and scheduling, between a safe-language-based microkernel and a hardware-isolation-based microkernel. The results correspond to \autoref{tab:seL4bench} The expected result is that safeL4 achieves better geometric-mean performance than seL4 across the evaluated microbenchmarks. This advantage primarily comes from avoiding privilege-level and address-space switches.

\subsubsection{Macrobenchmark Comparison with seL4 and Linux}

This experiment compares safeL4, seL4 LibOS, and Linux under realistic workloads by running YCSB on SQLite3. The YCSB benchmark is included in each corresponding system image and can be run directly. The results correspond to \autoref{fig:YCSB}. The expected overall trend is that safeL4 outperforms both seL4 and Linux.

\subsubsection{Microbenchmark Comparison with Redox}

This experiment compares safeL4 with Redox using file-system-related microbenchmarks to assess the performance impact of their different microkernel architectures. The results correspond to \autoref{tab:redox_comparison}. The expected overall trend is that safeL4 outperforms Redox. In safeL4, Redox's user-space RedoxFS and SD driver are ported into protection domains running in kernel mode, reducing IPC overhead and eliminating the costs of privilege-level and address-space switches.

%% file: sample.bib
@article{liedtke1996toward,
  title={Toward real microkernels},
  author={Liedtke, Jochen},
  journal={Communications of the ACM},
  volume={39},
  number={9},
  pages={70--77},
  year={1996},
  doi={10.1145/234215.234473},
  publisher={{ACM}}
}

@mastersthesis{herder2005towards,
  author={Herder, Jorrit N},
  title={Towards a true microkernel operating system},
  school={Vrije Universiteit Amsterdam},
  year={2005},
  url={https://minix3.org/doc/herder_thesis.pdf}
}

@inproceedings{sel42013,
  title={From {L3} to {seL4} what have we learnt in 20 years of {L4} microkernels?},
  author={Elphinstone, Kevin and Heiser, Gernot},
  booktitle={Proceedings of the 24th ACM Symposium on Operating Systems Principles (SOSP)},
  pages={133--150},
  year={2013},
  publisher={{ACM}},
  doi={10.1145/2517349.2522720}
}

@inproceedings{lied95,
  title={On micro-kernel construction},
  author={Liedtke, Jochen},
  booktitle={Proceedings of the 15th ACM Symposium on Operating Systems Principles (SOSP)},
  pages={237--250},
  doi={10.1145/224056.224075},
  publisher={{ACM}},
  year={1995}
}

@inproceedings{bomberger1992keykos,
  title={The {KeyKOS}\textregistered{} nanokernel architecture},
  author={Bomberger, Alan C and Hardy, Norman and Frantz, A Peri and Landau, Charles R and Frantz, William S and Shapiro, Jonathan S and Hardy, Ann C},
  booktitle={Proceedings of the USENIX Workshop on Micro-kernels and Other Kernel Architectures},
  pages={95--112},
  publisher={{USENIX} Association},
  url={https://flint.cs.yale.edu/cs428/doc/NanoKernel.pdf},
  year={1992}
}

@inproceedings{lackorzynski2009taming,
  title={Taming subsystems: Capabilities as universal resource access control in {L4}},
  author={Lackorzynski, Adam and Warg, Alexander},
  booktitle={Proceedings of the 2nd Workshop on Isolation and Integration in Embedded Systems (IIES)},
  pages={25--30},
  publisher={{ACM}},
  doi={10.1145/1519130.1519135},
  year={2009}
}

@inproceedings{shapiro1999eros,
  title={{EROS}: A fast capability system},
  author={Shapiro, Jonathan S and Smith, Jonathan M and Farber, David J},
  booktitle={Proceedings of the 17th ACM Symposium on Operating Systems Principles (SOSP)},
  pages={170--185},
  publisher={{ACM}},
  doi={10.1145/319151.319163},
  year={1999}
}

@article{polp,
  title={The protection of information in computer systems},
  author={Saltzer, Jerome H and Schroeder, Michael D},
  journal={Proceedings of the IEEE},
  volume={63},
  number={9},
  pages={1278--1308},
  year={1975},
  doi={10.1109/PROC.1975.9939},
  publisher={{IEEE}}
}

@inproceedings{jero2021practical,
  title={Practical principle of least privilege for secure embedded systems},
  author={Jero, Samuel and Furgala, Juliana and Pan, Runyu and Gadepalli, Phani Kishore and Clifford, Alexandra and Ye, Bite and Khazan, Roger and Ward, Bryan C and Parmer, Gabriel and Skowyra, Richard},
  booktitle={Proceedings of the 27th IEEE Real-Time and Embedded Technology and Applications Symposium (RTAS)},
  pages={1--13},
  year={2021},
  publisher={{IEEE}},
  doi={10.1109/RTAS52030.2021.00009}
}

@inproceedings{murray2013sel4,
  title={{seL4}: From general purpose to a proof of information flow enforcement},
  author={Murray, Toby and Matichuk, Daniel and Brassil, Matthew and Gammie, Peter and Bourke, Timothy and Seefried, Sean and Lewis, Corey and Gao, Xin and Klein, Gerwin},
  booktitle={Proceedings of the 34th IEEE Symposium on Security and Privacy (SP)},
  pages={415--429},
  year={2013},
  publisher={{IEEE}},
  doi={10.1109/SP.2013.35}
}

@inproceedings{sewell2011sel4,
  title={{seL4} enforces integrity},
  author={Sewell, Thomas and Winwood, Simon and Gammie, Peter and Murray, Toby and Andronick, June and Klein, Gerwin},
  booktitle={Proceedings of the 2nd International Conference on Interactive Theorem Proving (ITP)},
  pages={325--340},
  year={2011},
  publisher={Springer},
  doi={10.1007/978-3-642-22863-6\_24}
}

@inproceedings{brandauer2015parallel,
  title={Parallel objects for multicores: A glimpse at the parallel language {Encore}},
  author={Brandauer, Stephan and Castegren, Elias and Clarke, Dave and Fernandez-Reyes, Kiko and Johnsen, Einar Broch and Pun, Ka I and Tarifa, S Lizeth Tapia and Wrigstad, Tobias and Yang, Albert Mingkun},
  booktitle={Proceedings of the 15th International School on Formal Methods for the Design of Computer, Communication and Software Systems (SFM)},
  pages={1--56},
  year={2015},
  doi={10.1007/978-3-319-18941-3\_1},
  publisher={Springer}
}

@inproceedings{clebsch2015deny,
  title={Deny capabilities for safe, fast actors},
  author={Clebsch, Sylvan and Drossopoulou, Sophia and Blessing, Sebastian and McNeil, Andy},
  booktitle={Proceedings of the 5th International Workshop on Programming Based on Actors, Agents, and Decentralized Control (AGERE!)},
  pages={1--12},
  publisher={{ACM}},
  doi={10.1145/2824815.2824816},
  year={2015}
}

@article{klein2014comprehensive,
  title={Comprehensive formal verification of an {OS} microkernel},
  author={Klein, Gerwin and Andronick, June and Elphinstone, Kevin and Murray, Toby and Sewell, Thomas and Kolanski, Rafal and Heiser, Gernot},
  journal={ACM Transactions on Computer Systems (TOCS)},
  volume={32},
  number={1},
  pages={2:1--2:70},
  year={2014},
  doi={10.1145/2560537},
  publisher={{ACM}}
}

@inproceedings{accetta1986mach,
  title={Mach: A new kernel foundation for {UNIX} development},
  author={Accetta, Mike and Baron, Robert and Bolosky, William and Golub, David and Rashid, Richard and Tevanian, Avadis and Young, Michael},
  booktitle={Proceedings of the 1986 USENIX Summer Conference},
  pages={93--113},
  publisher={{USENIX} Association},
  url={https://eecs582.github.io/readings/mach-usenix86.pdf},
  year={1986}
}

@misc{sel4nio,
  title={{seL4} in software-defined vehicles: Vision, roadmap, and impact at {NIO}}, 
  author={Qu, Ning},
  howpublished={\url{https://sel4.systems/Summit/2024/slides/software-defined.pdf}},
  year={2024},
  note={Keynote at the {seL4} Summit 2024}
}

@inproceedings{hm2024,
  title={Microkernel goes general: Performance and compatibility in the {HongMeng} production microkernel},
  author={Chen, Haibo and Miao, Xie and Jia, Ning and Wang, Nan and Li, Yu and Liu, Nian and Liu, Yutao and Wang, Fei and Huang, Qiang and Li, Kun and Yang, Hongyang and Wang, Hui and Yin, Jie and Peng, Yu and Xu, Fengwei},
  booktitle={Proceedings of the 18th USENIX Symposium on Operating Systems Design and Implementation (OSDI)},
  pages={465--485},
  publisher={{USENIX} Association},
  url={https://www.usenix.org/system/files/osdi24-chen-haibo.pdf},
  year={2024}
}

@inproceedings{weinhold2023towards,
  title={Towards modular trusted execution environments},
  author={Weinhold, Carsten and Asmussen, Nils and G{\"o}hringer, Diana and Roitzsch, Michael},
  booktitle={Proceedings of the 6th Workshop on System Software for Trusted Execution (SysTEX)},
  pages={10--16},
  publisher={{ACM}},
  doi={10.1145/3578359.3593037},
  year={2023}
}

@inproceedings{miemietz2024perfect,
  title={A perfect fit? - Towards containers on microkernels},
  author={Miemietz, Till and Reusch, Viktor and Hille, Matthias and Kurze, Max and Lackorzynksi, Adam and Roitzsch, Michael and H{\"a}rtig, Hermann},
  booktitle={Proceedings of the 10th International Workshop on Container Technologies and Container Clouds (WoC)},
  pages={1--6},
  publisher={{ACM}},
  doi={10.1145/3702637.3702957},
  year={2024}
}

@inproceedings{miemietz2025metteagle,
  title={{MettEagle}: Costs and benefits of implementing containers on microkernels},
  author={Miemietz, Till and Reusch, Viktor and Hille, Matthias and Wrenger, Lars and Eisoldt, Jana and Kl{\"o}tzke, Jan and Kurze, Max and Lackorzynski, Adam and Roitzsch, Michael and H{\"a}rtig, Hermann},
  booktitle={Proceedings of the 19th USENIX Symposium on Operating Systems Design and Implementation (OSDI)},
  pages={979--996},
  publisher={{USENIX} Association},
  url={https://www.usenix.org/system/files/osdi25-miemietz.pdf},
  year={2025}
}

@inproceedings{shi2025serverless,
  title={Serverless functions made confidential and efficient with split containers},
  author={Shi, Jiacheng and Gu, Jinyu and Xia, Yubin and Chen, Haibo},
  booktitle={Proceedings of the 34th USENIX Security Symposium (USENIX Security)},
  pages={1091--1110},
  publisher={{USENIX} Association},
  url={https://www.usenix.org/system/files/usenixsecurity25-shi-jiacheng.pdf},
  year={2025}
}

@misc{diffmk08,
  author={Heiser, Gernot},
  title={What is the difference between a microkernel?},
  howpublished={\url{https://microkerneldude.org/2008/03/13/q-what-is-the-difference-between-a-microkernel}},
  year={2008}
}

@inproceedings{bershad1994spin,
  title={{SPIN}: An extensible microkernel for application-specific operating system services},
  author={Bershad, Brian N and Chambers, Craig and Eggers, Susan and Maeda, Chris and McNamee, Dylan and Pardyak, Przemyslaw and Savage, Stefan and Sirer, Emin G{\"u}n},
  booktitle={Proceedings of the 6th ACM SIGOPS European Workshop: Matching Operating Systems to Application Needs (EW)},
  pages={68--71},
  publisher={{ACM}},
  doi={10.1145/504390.504408},
  year={1994}
}

@inproceedings{spin1995,
  title={Extensibility, safety and performance in the {SPIN} operating system},
  author={Bershad, Brian N and Savage, Stefan and Pardyak, Przemyslaw and Sirer, Emin G{\"u}n and Fiuczynski, Marc E and Becker, David and Chambers, Craig and Eggers, Susan},
  booktitle={Proceedings of the 15th ACM Symposium on Operating Systems Principles (SOSP)},
  pages={267--284},
  publisher={{ACM}},
  doi={10.1145/224056.224077},
  year={1995}
}

@inproceedings{emmerich2019case,
  title={The case for writing network drivers in high-level programming languages},
  author={Emmerich, Paul and Ellmann, Simon and Bonk, Fabian and Egger, Alex and S{\'a}nchez-Torija, Esa{\'u} Garc{\'\i}a and G{\"u}nzel, Thomas and Di Luzio, Sebastian and Obada, Alexandru and Stadlmeier, Maximilian and Voit, Sebastian and Carle, Georg},
  booktitle={Proceedings of the 15th ACM/IEEE Symposium on Architectures for Networking and Communications Systems (ANCS)},
  pages={1--13},
  year={2019},
  doi={10.1109/ANCS.2019.8901892},
  publisher={{IEEE}}
}

@inproceedings{tock17,
  title={Multiprogramming a 64kB computer safely and efficiently},
  author={Levy, Amit and Campbell, Bradford and Ghena, Branden and Giffin, Daniel B and Pannuto, Pat and Dutta, Prabal and Levis, Philip},
  booktitle={Proceedings of the 26th Symposium on Operating Systems Principles (SOSP)},
  pages={234--251},
  publisher={{ACM}},
  doi={10.1145/3132747.3132786},
  year={2017}
}

@inproceedings{redleaf20,
  title={{RedLeaf}: Isolation and communication in a safe operating system},
  author={Narayanan, Vikram and Huang, Tianjiao and Detweiler, David and Appel, Dan and Li, Zhaofeng and Zellweger, Gerd and Burtsev, Anton},
  booktitle={Proceedings of the 14th USENIX Symposium on Operating Systems Design and Implementation (OSDI)},
  pages={21--39},
  publisher={{USENIX} Association},
  url={https://www.usenix.org/system/files/osdi20-narayanan_vikram.pdf},
  year={2020}
}

@inproceedings{theseus20,
  title={Theseus: An experiment in operating system structure and state management},
  author={Boos, Kevin and Liyanage, Namitha and Ijaz, Ramla and Zhong, Lin},
  booktitle={Proceedings of the 14th USENIX Symposium on Operating Systems Design and Implementation (OSDI)},
  pages={1--19},
  publisher={{USENIX} Association},
  url={https://www.usenix.org/system/files/osdi20-boos.pdf},
  year={2020}
}

@inproceedings{peng2025asterinas,
  title={{Asterinas}: A {Linux} {ABI}-compatible, {Rust}-based framekernel {OS} with a small and sound {TCB}},
  author={Peng, Yuke and Tian, Hongliang and Zhang, Junyang and Li, Ruihan and Chen, Chengjun and Jiang, Jianfeng and Xian, Jinyi and Wang, Xiaolin and Xu, Chenren and Zhou, Diyu and Luo, Yingwei and Yan, Shoumeng and Zhang, Yinqian},
  booktitle={Proceedings of the 2025 USENIX Annual Technical Conference (USENIX ATC)},
  pages={307--323},
  publisher={{USENIX} Association},
  url={https://www.usenix.org/system/files/atc25-peng-yuke.pdf},
  year={2025}
}

@article{modula3,
  title={Modula-3 language definition},
  author={Cardelli, Luca and Donahue, James and Glassman, Lucille and Jordan, Mick and Kalsow, Bill and Nelson, Greg},
  journal={ACM SIGPLAN Notices},
  volume={27},
  number={8},
  pages={15--42},
  year={1992},
  doi={10.1145/142137.142141},
  publisher={{ACM}}
}

@techreport{hunt2005overview,
  title={An overview of the {Singularity} project},
  author={Hunt, Galen and Larus, James and Abadi, Mart{\i}n and Aiken, Mark and Barham, Paul and F\"{a}hndrich, Manuel and Hawblitzel, Chris and Hodson, Orion and Levi, Steven and Murphy, Nick and Steensgaard, Bjarne and Tarditi, David and Wobber, Ted and Zill, Brian},
  number={MSR-TR-2005-135},
  year={2005},
  institution={Microsoft Research},
  url={https://www.microsoft.com/en-us/research/wp-content/uploads/2005/10/tr-2005-135.pdf},
  type={Technical Report}
}

@inproceedings{fahndrich2006language,
  title={Language support for fast and reliable message-based communication in {Singularity} {OS}},
  author={F{\"a}hndrich, Manuel and Aiken, Mark and Hawblitzel, Chris and Hodson, Orion and Hunt, Galen and Larus, James R and Levi, Steven},
  booktitle={Proceedings of the 1st ACM SIGOPS/EuroSys European Conference on Computer Systems (EuroSys)},
  pages={177--190},
  publisher={{ACM}},
  doi={10.1145/1217935.1217953},
  year={2006}
}

@article{singularity07,
  title={Singularity: Rethinking the software stack},
  author={Hunt, Galen C and Larus, James R},
  journal={ACM SIGOPS Operating Systems Review},
  volume={41},
  number={2},
  pages={37--49},
  year={2007},
  doi={10.1145/1243418.1243424},
  publisher={{ACM}}
}

@inproceedings{bricker1991new,
  title={A new look at microkernel-based {UNIX} operating systems: Lessons in performance and compatibility},
  author={Bricker, Allan and Gien, Michel and Guillemont, Marc and Lipkis, Jim and Orr, Douglas and Rozier, Marc},
  booktitle={Proceedings of the EurOpen Spring 1991 Conference},
  url={http://fqrd.fr/publications/chorus/CS-TR-91-7.pdf},
  year={1991}
}

@misc{sel4cantripos,
  author={Tate-Gans, June and Leffler, Sam and Yick, Kai},
  title={{CantripOS}: An {OS} for {Ambient ML} applications}, 
  year={2023},
  note={Keynote at the {seL4} Summit 2023},
  howpublished={\url{https://sel4.systems/Foundation/Summit/2023/slides/day2-01-leffler.pdf}},
}

@techreport{heiser2020sel4,
  title={The {seL4} microkernel – An introduction},
  author={Heiser, Gernot},
  institution={{seL4 Foundation}},
  year={2025},
  url={https://sel4.systems/About/seL4-whitepaper.pdf},
  type={White paper}
}

@incollection{jkernel99,
  title={{J-Kernel}: A capability-based operating system for {Java}},
  author={von Eicken, Thorsten and Chang, Chi-Chao and Czajkowski, Grzegorz and Hawblitzel, Chris and Hu, Deyu and Spoonhower, Dan},
  booktitle={Secure Internet Programming: Security Issues for Mobile and Distributed Objects},
  pages={369--393},
  year={1999},
  doi={10.1007/3-540-48749-2\_17},
  publisher={Springer}
}

@inproceedings{kaffeos00,
  title={Processes in {KaffeOS}: Isolation, resource management, and sharing in {Java}},
  author={Back, Godmar and Hsieh, Wilson C and Lepreau, Jay},
  booktitle={Proceedings of the 4th Symposium on Operating Systems Design and Implementation (OSDI)},
  pages={333--346},
  publisher={{USENIX} Association},
  url={https://dl.acm.org/doi/epdf/10.5555/1251229.1251252},
  year={2000}
}

@techreport{golm2002java,
  title={A {Java} operating system as the foundation of a secure network operating system},
  author={Golm, Michael and Felser, Meik and Wawersich, Christian and Klein{\"o}der, J{\"u}rgen},
  institution={Department of Computer Sciences 4 (Distributed Systems and Operating Systems), University of Erlangen-Nuremberg},
  type={Technical Report},
  number={TR-I4-02-05},
  url={https://www4.informatik.uni-erlangen.de/DE/Projects/JX/publications/jx-sec.pdf},
  year={2002}
}

@misc{Kry10,
  author={{Kry10 Unlimited, Inc.}},
  title={Kry10 - Enabling innovation in high trust systems},
  howpublished={\url{https://www.kry10.com/}},
  year={2024}
}

@misc{magnetite,
  author={Furgala, Juliana},
  title={Magnetite: {Rust}-based {OS} services for {seL4}}, 
  year={2023},
  note={Presentation slides at the {seL4} Summit 2023},
  howpublished={\url{https://sel4.systems/Foundation/Summit/2023/slides/day2-02-furgala.pdf}},
}

@misc{gapfruitos,
  author={{Gapfruit AG}},
  title={Gapfruit {OS}: Secure and scalable {OS}},
  howpublished={\url{https://www.gapfruit.com/}},
  year={2025}
}

@misc{sel4microkit,
  author={{seL4 Foundation}},
  title={The {seL4} {Microkit}},
  howpublished={\url{https://docs.sel4.systems/projects/microkit/}},
  year={2024}
}

@misc{camkesvm,
  author={{seL4 Foundation}},
  title={{CAmkES VMM}},
  howpublished={\url{https://docs.sel4.systems/projects/camkes-vm/}},
  year={2025}
}

@misc{lionsos,
  title={The {Lions Operating System} ({LionsOS})},
  author={{Trustworthy Systems Research Group, UNSW Sydney}},
  year={2024},
  howpublished={\url{https://lionsos.org}}
}

@article{dennis1966programming,
  title={Programming semantics for multiprogrammed computations},
  author={Dennis, Jack B and Van Horn, Earl C},
  journal={Communications of the ACM},
  volume={9},
  number={3},
  pages={143--155},
  year={1966},
  doi={10.1145/365230.365252},
  publisher={{ACM}}
}

@inproceedings{boyland2001cap,
  title={Capabilities for sharing: A generalisation of uniqueness and read-only},
  author={Boyland, John and Noble, James and Retert, William},
  booktitle={Proceedings of the 15th European Conference on Object-Oriented Programming (ECOOP)},
  pages={2--27},
  year={2001},
  doi={10.1007/3-540-45337-7\_2},
  publisher={Springer}
}

@inproceedings{haller2010cap,
  title={Capabilities for uniqueness and borrowing},
  author={Haller, Philipp and Odersky, Martin},
  booktitle={Proceedings of the 24th European Conference on Object-Oriented Programming (ECOOP)},
  pages={354--378},
  year={2010},
  doi={10.1007/978-3-642-14107-2\_17},
  publisher={Springer}
}

@inproceedings{li2021understanding,
  title={Understanding the overheads of hardware and language-based {IPC} mechanisms},
  author={Li, Zhaofeng and Huang, Tianjiao and Narayanan, Vikram and Burtsev, Anton},
  booktitle={Proceedings of the 11th Workshop on Programming Languages and Operating Systems (PLOS)},
  pages={53--61},
  publisher={{ACM}},
  doi={10.1145/3477113.3487275},
  year={2021}
}

@inproceedings{burtsev2021isolation,
  title={Isolation in {Rust}: What is missing?},
  author={Burtsev, Anton and Appel, Dan and Detweiler, David and Huang, Tianjiao and Li, Zhaofeng and Narayanan, Vikram and Zellweger, Gerd},
  booktitle={Proceedings of the 11th Workshop on Programming Languages and Operating Systems (PLOS)},
  pages={76--83},
  publisher={{ACM}},
  doi={10.1145/3477113.3487272},
  year={2021}
}

@inproceedings{du2019xpc,
  title={{XPC}: Architectural support for secure and efficient cross process call},
  author={Du, Dong and Hua, Zhichao and Xia, Yubin and Zang, Binyu and Chen, Haibo},
  booktitle={Proceedings of the 46th International Symposium on Computer Architecture (ISCA)},
  pages={671--684},
  publisher={{ACM}},
  doi={10.1145/3307650.3322218},
  year={2019}
}

@inproceedings{lampson1974protection,
  author={Lampson, Butler W.},
  title={Protection},
  booktitle={Proceedings of the 5th Princeton Symposium on Information Sciences and Systems},
  year={1971},
  pages={437--443},
  doi={10.1145/775265.775268},
  note={Reprinted in \textit{Operating Systems Review}, 8(1):18--24, 1974}
}

@article{lee2013preemptible,
  title={Preemptible {I/O} scheduling of garbage collection for solid state drives},
  author={Lee, Junghee and Kim, Youngjae and Shipman, Galen M and Oral, Sarp and Kim, Jongman},
  journal={IEEE Transactions on Computer-Aided Design of Integrated Circuits and Systems},
  volume={32},
  number={2},
  pages={247--260},
  year={2013},
  doi={10.1109/TCAD.2012.2227479},
  publisher={{IEEE}}
}

@inproceedings{tock25,
  title={Tock: From research to securing 10 million computers},
  author={Schuermann, Leon and Campbell, Brad and Ghena, Branden and Levis, Philip and Levy, Amit and Pannuto, Pat},
  booktitle={Proceedings of the ACM SIGOPS 31st Symposium on Operating Systems Principles (SOSP)},
  pages={36--49},
  publisher={{ACM}},
  doi={10.1145/3731569.3764828},
  year={2025}
}

@inproceedings{tg2009,
  author={Boyton, Andrew},
  title={A verified shared capability model},
  booktitle={Proceedings of the 4th International Workshop on Systems Software Verification (SSV)},
  pages={25--44},
  publisher={Elsevier},
  year={2009},
  doi={10.1016/J.ENTCS.2009.09.058},
}

@inproceedings{tg2008,
  title={Verified protection model of the {seL4} microkernel},
  author={Elkaduwe, Dhammika and Klein, Gerwin and Elphinstone, Kevin},
  booktitle={Proceedings of the 2nd International Conference on Verified Software: Theories, Tools, Experiments (VSTTE)},
  pages={99--114},
  year={2008},
  doi={10.1007/978-3-540-87873-5\_11},
  publisher={Springer}
}

@mastersthesis{objcapcom2019,
    author={Hillert, Jessica},
    title={A comparison of the capability systems of {Encore}, {Pony} and {Rust}},
    school={Uppsala University},
    year={2019},
    type={Bachelor's thesis},
    url={https://uu.diva-portal.org/smash/get/diva2:1363822/FULLTEXT01.pdf}
}

@book{levy2014capability,
    title={Capability-based computer systems},
    author={Levy, Henry M},
    year={1984},
    doi={10.1016/C2013-0-01290-X},
    publisher={Digital Press}
}

@inproceedings{javaobj10,
  title={{Joe-E}: A security-oriented subset of {Java}},
  author={Mettler, Adrian and Wagner, David A and Close, Tyler},
  booktitle={Proceedings of the 17th Network and Distributed System Security Symposium (NDSS)},
  pages={357--374},
  publisher={The Internet Society},
  url={https://www.ndss-symposium.org/wp-content/uploads/2017/09/met.pdf},
  year={2010}
}

@phdthesis{eobj06,
  title={Robust composition: Towards a unified approach to access control and concurrency control},
  author={Miller, Mark Samuel},
  year={2006},
  school={Johns Hopkins University},
  url={https://jscholarship.library.jhu.edu/server/api/core/bitstreams/c2c3ea50-2613-47a7-b33c-9383c706ee63/content}
}

@inproceedings{objcap2000,
  title={Capability-based financial instruments},
  author={Miller, Mark S and Morningstar, Chip and Frantz, Bill},
  booktitle={Proceedings of the 4th International Conference on Financial Cryptography (FC)},
  pages={349--378},
  year={2000},
  doi={10.1007/3-540-45472-1\_24},
  publisher={Springer}
}

@techreport{emily2006,
  author={Stiegler, Marc and Miller, Mark},
  title={How {Emily} tamed the {Caml}},
  institution={Hewlett-Packard Laboratories},
  year={2006},
  url={https://shiftleft.com/mirrors/www.hpl.hp.com/techreports/2006/HPL-2006-116.pdf},
  type={Technical Report}
}

@article{objcap2017,
  title={Robust and compositional verification of object capability patterns},
  author={Swasey, David and Garg, Deepak and Dreyer, Derek},
  journal={Proceedings of the ACM on Programming Languages},
  volume={1},
  number={OOPSLA},
  pages={89:1--89:26},
  year={2017},
  doi={10.1145/3133913},
  publisher={{ACM}}
}

@article{javaobj24,
  title={Extending the object-capability model with fine-grained type-based capabilities},
  author={Wism{\"u}ller, Roland and Ludwig, Damian and Breitweiser, Felix},
  journal={Journal of Object Technology},
  volume={23},
  number={1},
  pages={1-36},
  doi={10.5381/JOT.2024.23.1.A1},
  year={2024}
}

@inproceedings{statespill,
  title={A characterization of state spill in modern operating systems},
  author={Boos, Kevin and Vecchio, Emilio Del and Zhong, Lin},
  booktitle={Proceedings of the 12th European Conference on Computer Systems (EuroSys)},
  pages={389--404},
  publisher={{ACM}},
  doi={10.1145/3064176.3064205},
  year={2017}
}

@inproceedings{peng2024framekernel,
  title={Framekernel: A safe and efficient kernel architecture via {Rust}-based intra-kernel privilege separation},
  author={Peng, Yuke and Tian, Hongliang and Xian, Jinyi and Zhou, Shuai and Yan, Shoumeng and Zhang, Yinqian},
  booktitle={Proceedings of the 15th ACM SIGOPS Asia-Pacific Workshop on Systems (APSys)},
  pages={31--37},
  publisher={{ACM}},
  doi={10.1145/3678015.3680492},
  year={2024}
}

@article{rustbelt2017,
  title={{RustBelt}: Securing the foundations of the {Rust} programming language},
  author={Jung, Ralf and Jourdan, Jacques-Henri and Krebbers, Robbert and Dreyer, Derek},
  journal={Proceedings of the ACM on Programming Languages},
  volume={2},
  number={POPL},
  pages={66:1--66:34},
  year={2018},
  doi={10.1145/3158154},
  publisher={{ACM}}
}

@misc{Unwinding,
  author={Guo, Gary},
  title={{Unwinding}: Unwinding library in {Rust} and for {Rust}},
  year={2024},
  howpublished={\url{https://github.com/nbdd0121/unwinding}}
}

@techreport{bonwick2003zettabyte,
  title={The {Zettabyte} file system},
  author={Bonwick, Jeff and Ahrens, Matt and Henson, Val and Maybee, Mark and Shellenbaum, Mark},
  year={2003},
  institution={{Sun Microsystems, Inc.}},
  url={https://web.mit.edu/6.033/www/papers/zfs.pdf},
  type={Technical Report},
}

@misc{redox,
  author={{The {Redox} Developers}},
  title={{Redox OS}},
  year={2025},
  howpublished={\url{https://www.redox-os.org}}
}

@misc{redoxfs,
  author={{The {Redox} Developers}},
  title={{RedoxFS}: The {Redox} filesystem},
  year={2018},
  howpublished={\url{https://gitlab.redox-os.org/redox-os/redoxfs}}
}

@misc{pod,
  author={{The cppreference.com contributors}},
  title={{C}++ named requirements: {PODType} (deprecated in {C}++20)},
  year={2022},
  howpublished={\url{https://en.cppreference.com/w/cpp/named_req/PODType}}
}

@article{harrison1976protection,
  title={Protection in operating systems},
  author={Harrison, Michael A and Ruzzo, Walter L and Ullman, Jeffrey D},
  journal={Communications of the ACM},
  volume={19},
  number={8},
  pages={461--471},
  year={1976},
  doi={10.1145/360303.360333},
  publisher={{ACM}}
}

@article{tg1977,
  title={A linear time algorithm for deciding subject security},
  author={Lipton, Richard J and Snyder, Lawrence},
  journal={Journal of the ACM},
  volume={24},
  number={3},
  pages={455--464},
  year={1977},
  doi={10.1145/322017.322025},
  publisher={{ACM}}
}

@inproceedings{ycsb,
  title={Benchmarking cloud serving systems with {YCSB}},
  author={Cooper, Brian F and Silberstein, Adam and Tam, Erwin and Ramakrishnan, Raghu and Sears, Russell},
  booktitle={Proceedings of the 1st ACM Symposium on Cloud Computing (SoCC)},
  pages={143--154},
  publisher={{ACM}},
  doi={10.1145/1807128.1807152},
  year={2010}
}

@misc{sqlite,
  author={{SQLite Consortium}},
  title={{SQLite}},
  year={2025},
  howpublished={\url{https://www.sqlite.org/index.html}}
}

@inproceedings{lmbench,
  title={{lmbench}: Portable tools for performance analysis},
  author={McVoy, Larry W and Staelin, Carl},
  booktitle={Proceedings of the 1996 USENIX Annual Technical Conference (USENIX ATC)},
  pages={279--294},
  year={1996},
  publisher={{USENIX} Association}
}

@misc{sel4bench,
  author={{seL4 Foundation}},
  title={The sel4bench suite},
  year={2025},
  howpublished={\url{https://docs.sel4.systems/projects/sel4bench}}
}

@article{hardy1988confused,
  title={The confused deputy (or why capabilities might have been invented)},
  author={Hardy, Norm},
  journal={ACM SIGOPS Operating Systems Review},
  volume={22},
  number={4},
  pages={36--38},
  year={1988},
  doi={10.1145/54289.871709},
  publisher={{ACM}}
}

@techreport{miller2003capability,
  title={Capability myths demolished},
  author={Miller, Mark S and Yee, Ka-Ping and Shapiro, Jonathan},
  number={SRL2003-02},
  year={2003},
  institution={Systems Research Laboratory, Johns Hopkins University},
  url={http://zesty.ca/capmyths/usenix.pdf},
  type={Technical Report}
}

@misc{l4re,
  author={{Kernkonzept GmbH}},
  title={The L4Re operating system framework},
  howpublished={\url{https://l4re.org}},
  year={2026}
}

@misc{rust_inline_assembly_rules,
  author={{The Rust Project Developers}},
  title={Rust rules for inline assembly},
  howpublished={\url{https://doc.rust-lang.org/reference/inline-assembly.html}},
  year={2026}
}

@inproceedings{potyondy2026rage,
  title={Rage against the state machine: Type-stated hardware peripherals for increased driver correctness},
  author={Potyondy, Tyler and Tarbinian, Anthony and Schuermann, Leon and Mugnier, Eric and Ackerman, Adin and Levy, Amit and Pannuto, Pat},
  booktitle={Proceedings of the 31st ACM International Conference on Architectural Support for Programming Languages and Operating Systems, Volume 2},
  pages={1508--1522},
  publisher={{ACM}},
  doi={10.1145/3779212.3790207},
  year={2026},
  series={ASPLOS}
}

@inproceedings{hallgren2005principled,
  title={A principled approach to operating system construction in {Haskell}},
  author={Hallgren, Thomas and Jones, Mark P and Leslie, Rebekah and Tolmach, Andrew},
  booktitle={Proceedings of the 10th ACM SIGPLAN International Conference on Functional Programming (ICFP)},
  pages={116--128},
  publisher={{ACM}},
  doi = {10.1145/1086365.1086380},
  year={2005}
}

@inproceedings{chen2025atmosphere,
  title={Atmosphere: Practical verified kernels with {Rust} and {Verus}},
  author={Chen, Xiangdong and Li, Zhaofeng and Zhang, Jerry and Narayanan, Vikram and Burtsev, Anton},
  booktitle={Proceedings of the ACM SIGOPS 31st Symposium on Operating Systems Principles (SOSP)},
  pages={752--767},
  year={2025},
  doi={10.1145/3731569.3764821},
  publisher={ACM}
}
